\documentclass[a4paper,12pt]{article}
\newcommand{\plei}[1]{ $\alpha'_{j} \sim \mathcal{N}$} 
\usepackage{tabularx}
\newcolumntype{R}{@{\extracolsep{1.5cm}}r@{\extracolsep{6pt}}} 
\usepackage{epsfig}
\usepackage{amsfonts}
\usepackage{url}
\usepackage{notoccite}
\usepackage{booktabs}
\usepackage{color}
\usepackage{graphicx}
\usepackage{footnote}
\usepackage{epsfig}
\usepackage{amsfonts}
\usepackage{amsmath}
\usepackage{amssymb}
\usepackage{rotating}
\usepackage{bibunits}
\usepackage{multirow}
\usepackage{mathtools}
\usepackage{chngpage}
\usepackage{setspace}
\usepackage{array}
\usepackage{paralist}
\usepackage{authblk}
\usepackage{setspace}
\usepackage[bottom]{footmisc}
\usepackage[singlelinecheck=false, footnotesize, skip=0pt]{caption} 
\DeclareMathOperator{\se}{se} 
\DeclareMathOperator{\cor}{cor}
\DeclareMathOperator{\cov}{cov}
\DeclareMathOperator{\var}{var}

\usepackage{enumitem}
\newcommand\independent{\protect\mathpalette{\protect\independenT}{\perp}}
\def\independenT#1#2{\mathrel{\rlap{$#1#2$}\mkern2mu{#1#2}}}
\begin{document}
\title{Extending the MR-Egger method for multivariable Mendelian randomization to correct for both measured and unmeasured pleiotropy}
\author{Jessica MB Rees\textsuperscript{1} \qquad Angela Wood\textsuperscript{1} \qquad Stephen Burgess\textsuperscript{1,2,\thanks{Address: MRC Biostatistics Unit, University of Cambridge, Cambridge, Cambridgeshire, CB2 0SR, UK. Correspondence to: sb452@medschl.cam.ac.uk.}}}
\affil{\textsuperscript{1} \small{Cardiovascular Epidemiology Unit, University of Cambridge, UK.}}
\affil{\textsuperscript{2}  \small{MRC Biostatistics Unit, University of Cambridge, Cambridge, UK.}}
\maketitle
\noindent \textbf{Short title:} Multivariable MR-Egger method for Mendelian randomization.
\setstretch{1.2}
\pagebreak
\begin{bibunit}[wileyj]
\section*{Abstract}
Methods have been developed for Mendelian randomization that can obtain consistent causal estimates while relaxing the instrumental variable assumptions. These include multivariable Mendelian randomization, in which a genetic variant may be associated with multiple risk factors so long as any association with the outcome is via the measured risk factors (measured pleiotropy), and the MR-Egger (Mendelian randomization-Egger) method, in which a genetic variant may be directly associated with the outcome not via the risk factor of interest, so long as the direct effects of the variants on the outcome are uncorrelated with their associations with the risk factor (unmeasured pleiotropy). In this paper, we extend the MR-Egger method to a multivariable setting to correct for both measured and unmeasured pleiotropy. We show, through theoretical arguments and a simulation study, that the multivariable MR-Egger method has advantages over its univariable counterpart in terms of plausibility of the assumption needed for consistent causal estimation, and power to detect a causal effect when this assumption is satisfied. The methods are compared in an applied analysis to investigate the causal effect of high-density lipoprotein cholesterol on coronary heart disease risk. The multivariable MR-Egger method will be useful to analyse high-dimensional data in situations where the risk factors are highly related and it is difficult to find genetic variants specifically associated with the risk factor of interest (multivariable by design), and as a sensitivity analysis when the genetic variants are known to have pleiotropic effects on measured risk factors.

\textbf{Keywords:} Mendelian randomization, invalid instruments, pleiotropy, MR-Egger, multivariable. 
\pagebreak

\section{Introduction}
Mendelian randomization (MR) uses genetic variants as instrumental variables to estimate the causal effect of a risk factor on an outcome using observational data \cite{smith2003, lawlor2008}. Increases in the scale of genome-wide association studies have led to large numbers of genetic variants that are associated with candidate risk factors being discovered \cite{sleiman2010}. If the variants explain additional variability in the risk factor then using multiple variants in a MR analysis will increase power to detect a causal effect \cite{brion2013power, freeman2013power}. A pleiotropic genetic variant is associated with multiple risk factors; such a variant is not a valid instrumental variable and its inclusion in a (univariable) MR analysis may result in biased causal estimates and inappropriate inferences \cite{burgess2013scores}. As more variants are used in an MR analysis, the chance of including a pleiotropic variant increases.

For some sets of risk factors, including lipid fractions, several risk factors have common genetic predictors. Although such genetic variants are pleiotropic, they can be used to estimate causal effects in a multivariable MR framework \cite{burgess2015multivariable}. In multivariable MR, the instrumental variable assumptions are extended to allow a genetic variant to be associated with multiple risk factors, provided all associated risk factors are included in the analysis. Alternatively, when genetic variants are suspected to violate the instrumental variable assumptions through unknown pleiotropic pathways, methods have been developed to estimate consistent causal effects under weaker assumptions. These include the weighted median \cite{bowden2016consistent} and MR-Egger \cite{bowden2015egger} methods. The extension of MR-Egger to a multivariable setting has been implemented by Helgadottir \textit{et al.}\
\cite{helgadottir2016} as part of a sensitivity analysis in their applied work investigating the effect of lipid fractions on coronary heart disease (CHD) risk. However, there remains several methodological issues relating to the implementation of the method, and the assumptions required.

In this paper, we expand univariable MR-Egger to the multivariable setting. In Section~\ref{sec:methods}, we introduce the conventional and MR-Egger methods in both univariable and multivariable contexts. We provide an example analysis using published data on lipid fractions and CHD risk (Section~\ref{sec:example}), and compare results from the different MR methods in a simulation study (Section~\ref{sec:simulation}). Finally (Section~\ref{sec:discussion}), we discuss the results of the paper and the implications for applied practice. Software code for implementing all of the methods used in this paper is provided in the Web Appendix.

\section{Methods}
\label{sec:methods}
Initially, we consider the causal effect of a risk factor $X$ on an outcome $Y$ using genetic variants $G_{j}$ ($j = 1, \ldots, J$) that are assumed to be uncorrelated (not in linkage disequilibrium). Then, we expand to consider multiple risk factors $X_1, X_2, \ldots, X_K$. Increasingly, MR investigations are implemented using summarized data from consortia to leverage their large sample sizes, thereby improving the precision of causal estimates \cite{burgess2015publishedata}. We therefore assume that summarized data are available on the associations of each genetic variant with the risk factor (or with each risk factor for the multivariable setting) and with the outcome: the beta-coefficients ($\hat{\beta}_{X_{j}}, \hat{\beta}_{Y_{j}}$) and their standard errors ($\se(\hat{\beta}_{X_{j}}), \se(\hat{\beta}_{Y_{j}})$) from univariable regression on each variant $G_{j}$ in turn. We additionally assume that the associations of genetic variants with the risk factor and the outcome, and the causal effect of the risk factor on the outcome, are linear and homogeneous across the population; these assumptions are discussed in detail elsewhere \cite{burgess2016multiple}. To distinguish between the parameters from the different methods considered, we use the following subscript notation: UI (`univariable inverse variance weighted (IVW)'); UE (`univariable MR-Egger'); MI (multivariable IVW); and ME (`multivariable MR-Egger').   

\subsection{Univariable Mendelian randomization}
In a univariable MR analysis, each genetic variant must satisfy the following criteria to be a valid instrumental variable (IV):  \\
\begin{compactitem}
\item IV1: the variant is associated with the risk factor $X$, 
\item IV2: the variant is independent of all confounders $U$ of the risk factor--outcome association, and
\item IV3: the variant is independent of the outcome $Y$ conditional on the risk factor $X$ and confounders $U$ \cite{greenland2000}. \\
\end{compactitem}

\begin{figure}[htb]
\centering
\includegraphics[scale=0.2]{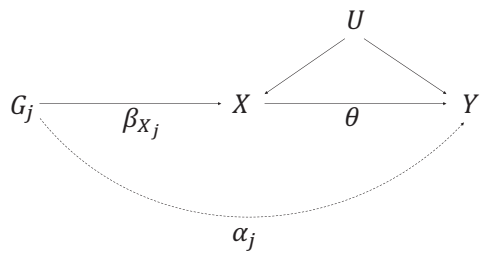}
\caption{Causal directed acyclic graph illustrating univariable Mendelian randomization assumptions with potential violation of IV3 by a pleiotropic effect indicated by a dotted line. The genetic effect of $G_{j}$ on $X$ is $\beta_{X_{j}}$, the direct (pleiotropic) effect of $G_{j}$ on $Y$ via an independent pathway is $\alpha_{j}$ (representing the potential violation of the IV3 assumption), and the causal effect of the risk factor $X$ on the outcome $Y$ is $\theta$. $U$ represents the set of variables that confound the association between $X$ and $Y$.} \label{dag1}
\end{figure}

These assumptions imply that the genetic variant should not have an effect on the outcome except via the risk factor. Under linearity assumptions, the association between the genetic variant and the outcome can be decomposed into an indirect effect via the risk factor and a direct effect:
\begin{equation}
\beta_{Yj} = \alpha_j + \theta \beta_{X_{j}}
\end{equation}
where $\theta$ is the causal effect of the risk factor on the outcome. Genetic variant $j$ is pleiotropic if $\alpha_j \neq 0$, and $\alpha_j$ is the direct effect of the genetic variant on the outcome. Figure~\ref{dag1} contains a direct effect $\alpha_j$ via an independent pathway, which violates the IV3 assumption.

\begin{center}
[Figure 1 should appear about here.]
\end{center}

With a single genetic variant, $G_1$ say, the causal estimate is $\hat{\beta}_{Y_{1}}/\hat{\beta}_{X_{1}}$ \cite{didelez2007}. This is a consistent estimate of the causal effect $\theta$ when $\alpha_1 = 0$. With multiple genetic variants, the inverse-variance weighted (IVW) estimate is the weighted average of these causal estimates \cite{johnson2013}, using the inverse of their approximate variances $\se(\hat{\beta}_{Y_{j}})^2/\hat{\beta}_{X_{j}}^2$ as weights:
\begin{equation}
\hat{\theta}_{UI} = \frac{\sum_j \hat{\beta}_{Y_{j}} \hat{\beta}_{X_{j}} \se(\hat{\beta}_{Y_{j}})^{-2}}{\sum_j \hat{\beta}_{X_{j}}^2 \se(\hat{\beta}_{Y_{j}})^{-2}} \label{eq:IVW_simple}
\end{equation}
This estimate can also be obtained from individual-level data using the two-stage least squares method \cite{angrist1995twostage}. Alternatively, the causal effect of the risk factor on the outcome can be estimated using a weighted linear regression of the genetic association estimates \cite{burgess2013ivw}, with the intercept set to zero:
\begin{equation}
\hat{\beta}_{Y_{j}} = \theta_{UI} \hat{\beta}_{X_{j}} + \epsilon_{UI_{j}}, \quad \mbox{weights = } \se(\hat{\beta}_{Y_{j}})^{-2} \label{eq:IVW}
\end{equation}
The above weighted regression model, where the residual standard error is set to one, is equivalent to performing a fixed-effect meta-analysis of the variant-specific causal estimates \cite{thompson1999}. Under a multiplicative random-effects model, the residual standard error can be greater than one, allowing for heterogeneity in the causal estimates.  The point estimate from the fixed- and random-effect models will be the same, but the standard error of the causal effect from the multiplicative random-effects model will be larger if there is heterogeneity between the causal estimates. Throughout this paper, we apply a multiplicative random-effects model to all the analyses.  

The MR-Egger estimate is obtained using the same regression model as equation~\ref{eq:IVW_simple}, but allowing the intercept to be estimated \cite{bowden2015egger}:
\begin{equation}
\hat{\beta}_{Y_{j}} = \theta_{0UE} + \theta_{UE} \hat{\beta}_{X_{j}} + \epsilon_{UE_{j}}, \quad \mbox{weights = } \se(\hat{\beta}_{Y_{j}})^{-2} \label{eq:uv_egger}
\end{equation}
If the genetic variants are not pleiotropic, then the intercept term should tend to zero as the sample size increases, and the MR-Egger estimate ($\hat{\theta}_{UE}$) and the IVW estimate ($\hat{\theta}_{UI}$) are both consistent estimates of the causal effect. Additionally, if the genetic variants are pleiotropic but the direct effects $\boldsymbol{\alpha}$ (bold symbols represent vectors across the $j$ genetic variants) are independent of the associations of the variants with the risk factor $\boldsymbol{\beta_{X}}$ (known as the InSIDE assumption -- Instrument Strength Independent of Direct Effect), then the MR-Egger estimate will be a consistent estimate of $\theta$ \cite{bowden2015egger,kolesar2014}. 

Under the InSIDE assumption, the intercept term $\hat{\theta}_{0UE}$ can be interpreted as an estimate of the average direct effect of the genetic variants \cite{bowden2016consistent}. If the average direct effect is zero (referred to as `balanced pleiotropy'), and the InSIDE assumption is satisfied, the intercept term should tend to zero as the sample size increases, and the MR-Egger estimate ($\hat{\theta}_{UE}$) and the IVW estimate ($\hat{\theta}_{UI}$) are both consistent estimates of the causal effect. If the intercept term differs from zero, then either the InSIDE assumption is violated or the average direct effect differs from zero (referred to as `directional pleiotropy'); this is a test of the validity of the instrumental variable assumptions (the MR-Egger intercept test). 

\subsection{Multivariable Mendelian randomization}
In a multivariable MR analysis, each genetic variant must satisfy the following criteria: \\
\begin{compactitem}
\item IV1(M): the variant is associated with at least one of the risk factors $X_{k}$,
\item IV2(M): the variant is independent of all confounders $U$ of each of the risk factor--outcome associations, and
\item IV3(M): the variant is independent of the outcome $Y$ conditional on the risk factors $X_{k}$ and confounders $U$ \cite{burgess2015multivariable}.  \\
\end{compactitem}

\begin{figure}[htb]
\centering
\includegraphics[scale=0.3]{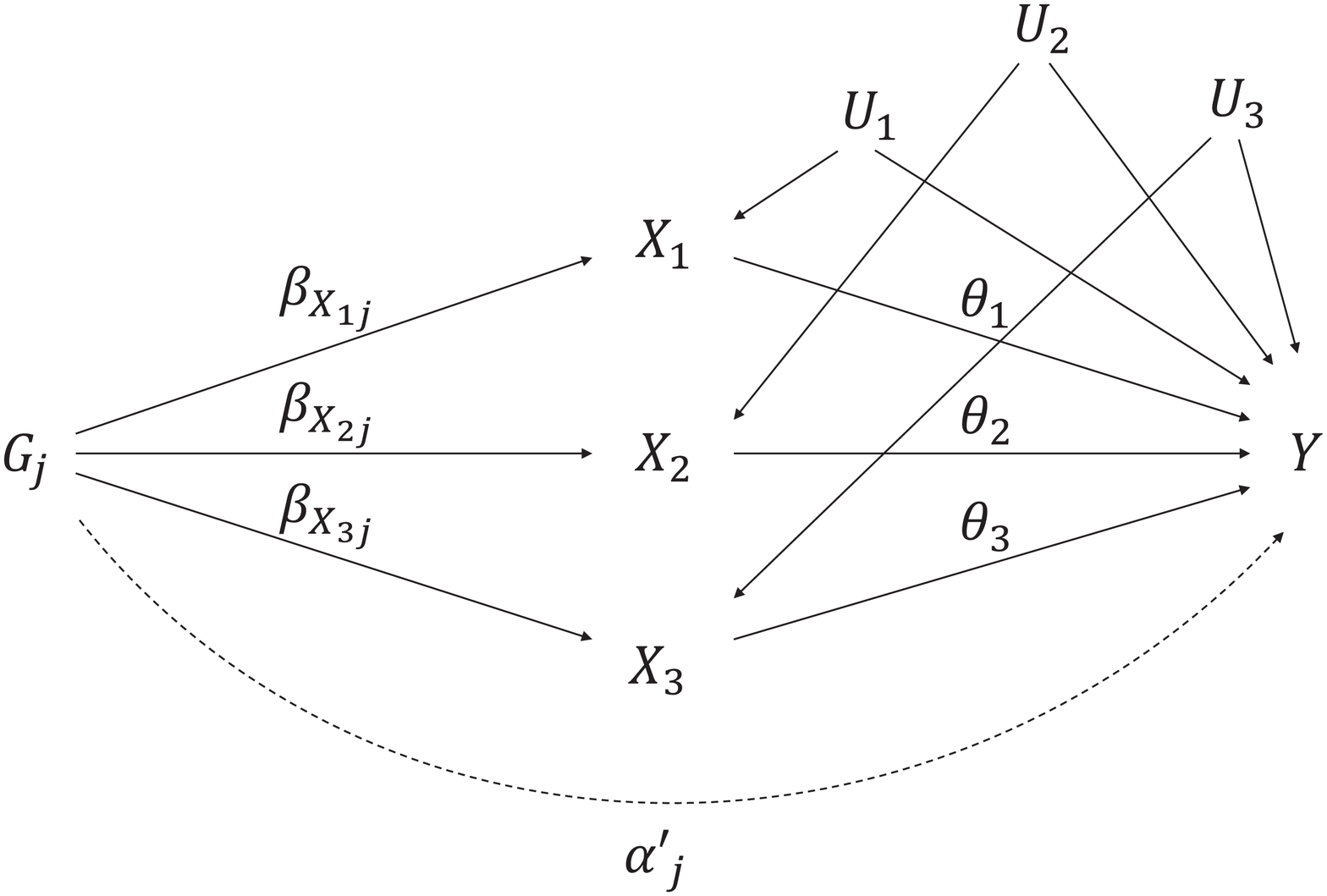}
\caption{Causal directed acyclic graph illustrating multivariable Mendelian randomization assumptions for a set of genetic variants $G_{j}$, three risk factors $X_{1}$, $X_{2}$ and $X_{3}$, and outcome Y.  The genetic effect of $G_{j}$ on $X_{k}$ is $\beta_{X_{kj}}$, the direct (pleiotropic) effect of $G_{j}$ on $Y$ is $\alpha'_{j}$, and the causal effect of the risk factor $X_{k}$ on the outcome $Y$ is $\theta_{k}$. $U_k$ represents the set of variables that confound the associations between $X_k$ and $Y$.} \label{dag2}
\end{figure}

Now, the association of the genetic variants with the outcome can be decomposed into indirect effects via each of the risk factors and a residual direct effect $\alpha'_j$. Assuming there are 3 risk factors and all relationships are linear:
\begin{equation}
\beta_{Y_{j}} = \alpha'_j + \theta_1 \beta_{X_{1j}} + \theta_2 \beta_{X_{2j}} + \theta_3 \beta_{X_{3j}}
\label{eq:multivariable}
\end{equation}
where $\theta_k$ is the causal effect of the risk factor $k$ on the outcome (Figure~\ref{dag2}). We assume that the risk factors do not have causal effects on each other; we later relax this assumption and allow for causal effects between the risk factors. 

\begin{center}
[Figure 2 should appear about here.]
\end{center}

As in the univariable setting, causal estimates of the effect of each risk factor on the outcome can be obtained from individual-level data using the two-stage least squares method \cite{burgess2015multivariable}. The same estimates can also be obtained using multivariable weighted linear regression of the genetic association estimates, with the intercept set to zero (referred to as the multivariable IVW method) \cite{burgess2015letter}:
\begin{equation}
\hat{\beta}_{Y_{j}} = \theta_{1MI} \hat{\beta}_{X_{1j}} + \theta_{2MI} \hat{\beta}_{X_{2j}} + \theta_{3MI} \hat{\beta}_{X_{3j}} + \epsilon_{MI_{j}}, \quad \mbox{weights = } \se(\hat{\beta}_{Y_{j}})^{-2} \label{eq:mvivw}
\end{equation}

We propose the natural extension to multivariable MR-Egger using the same regression model, but allowing the intercept to be estimated:
\begin{equation}
\hat{\beta}_{Y_{j}} = \theta_{0ME} + \theta_{1ME} \hat{\beta}_{X_{1j}} + \theta_{2ME} \hat{\beta}_{X_{2j}} + \theta_{3ME} \hat{\beta}_{X_{3j}} + \epsilon_{ME_{j}}, \quad \mbox{weights = } \se(\hat{\beta}_{Y_{j}})^{-2}
\label{eq:mvegg}
\end{equation}

\subsection{Assumptions for multivariable MR-Egger}
\label{sec:inside}
We assume that the causal effect of risk factor 1 ($\theta_1$) is of interest and provide the assumptions necessary for the MR-Egger estimate of $\theta_1$ to be consistent. If all of the causal effects are to be interpreted then these assumptions must apply for each risk factor.

If the $\boldsymbol{\beta_{X_{1}}}$ parameters are independent of the $\boldsymbol{\beta_{X_{k}}}$ parameters for all $k = 2, 3, \ldots, K$, then the InSIDE assumption for multivariable MR-Egger is satisfied if the direct effects of the genetic variants $\boldsymbol{\alpha'}$ are independent of $\boldsymbol{\beta_{X_{1}}}$.  More formally, we require: 
\begin{equation}
\boldsymbol{\beta_{X_{1}}} \independent \boldsymbol{\alpha'}, \quad \mbox{if }  \boldsymbol{\beta_{X_{1}}} \independent \boldsymbol{\beta_{X_{2}}}, \ldots, \boldsymbol{\beta_{X_{K}}}
\end{equation}
for the estimate of $\theta_1$ from multivariable MR-Egger to be consistent. If the InSIDE assumption is satisfied, then the weighted covariance of $\boldsymbol{\beta_{X_{1}}}$ and $\boldsymbol{\alpha'}$ ($\cov_w(\boldsymbol{\alpha'}, \boldsymbol{\beta_{X_{1}}}$)) will tend to zero as the number of genetic variants $J$ tends to infinity. The estimate of $\theta_1$ from multivariable MR-Egger when the $\boldsymbol{\beta_{X_{1}}}$ parameters are independent of $\boldsymbol{\beta_{X_{k}}}$ for all $k = 2, 3, \ldots, K$ is:
\begin{equation}
\hat{\theta}_{1ME} = \frac{\cov_w(\boldsymbol{\hat{\beta}_{Y}}, \boldsymbol{\hat{\beta}_{X_{1}}})}{\var_w(\boldsymbol{\hat{\beta}_{X_{1}}})} \xrightarrow{N \rightarrow \infty} \frac{\cov_w(\boldsymbol{\beta_{Y}}, \boldsymbol{\beta_{X_{1}}})}{\var_w(\boldsymbol{\beta_{X_{1}}})} = \theta_1 + \frac{\cov_w(\boldsymbol{\alpha'}, \boldsymbol{\beta_{X_{1}}})}{\var_w(\boldsymbol{\beta_{X_{1}}})}
\label{eq:insidein}
\end{equation}
which is equal to $\theta_1$ if the InSIDE assumption is satisfied, where $\cov_w$ and $\var_w$ represent the weighted covariance and weighted variance using the inverse-variance weights $\se(\hat{\beta}_{Yj})^{-2}$:

\begin{align}
\cov_w(\boldsymbol{\alpha'}, \boldsymbol{\beta_{X_{1}}}) &= \frac{\sum_j (\alpha'_j - \bar{\alpha'}_w) (\beta_{X_{1j}} - \bar{\beta}_{X_{1}w}) \se(\hat{\beta}_{Yj})^{-2}}{\sum_j \se(\hat{\beta}_{Yj})^{-2}} \notag \\
\var_w(\boldsymbol{\beta_{X_{1}}}) &= \frac{\sum_j (\beta_{X_{1j}} - \bar{\beta}_{X_{1}w})^2 \se(\hat{\beta}_{Yj})^{-2}}{\sum_j \se(\hat{\beta}_{Yj})^{-2}} \notag \\
\bar{\alpha'}_w        &= \frac{\sum_j \alpha'_j \se(\hat{\beta}_{Yj})^{-2}}{\sum_j \se(\hat{\beta}_{Yj})^{-2}} \notag \\
\bar{\beta}_{X_{1}w}       &= \frac{\sum_j \beta_{X_{1j}} \se(\hat{\beta}_{Yj})^{-2}}{\sum_j \se(\hat{\beta}_{Yj})^{-2}} 
\end{align}

If the $\boldsymbol{\beta_{X_{1}}}$ parameters are correlated with at least one of the sets of $\boldsymbol{\beta_{X_{k}}}$ parameters ($k = 2, 3, \ldots, K$), then the InSIDE assumption is required to hold for $\boldsymbol{\beta_{X_{1}}}$ and for all of the $\boldsymbol{\beta_{X_{k}}}$ parameters that are correlated with  $\boldsymbol{\beta_{X_{1}}}$. More formally, we require: 
\begin{equation}
\boldsymbol{\beta_{X_{k}}} \independent \boldsymbol{\alpha'}, \quad \mbox{for all } \boldsymbol{\beta_{X_{k}}} \mbox{correlated with } \boldsymbol{\beta_{X_{1}}} (\mbox{including } \boldsymbol{\beta_{X_{1}}} \mbox{itself})
\end{equation}
For example, if $k=2$, and $\boldsymbol{\beta_{X_{1}}}$ is correlated with $\boldsymbol{\beta_{X_{2}}}$, we require both of the weighted covariances of $\boldsymbol{\alpha'}$ with $\boldsymbol{\beta_{X_{1}}}$ and $\boldsymbol{\beta_{X_{2}}}$ to be zero to produce a consistent estimate of $\theta_1$. The estimate of $\theta_1$ from multivariable MR-Egger with two risk factors where $\boldsymbol{\beta_{X_{1}}}$ and $\boldsymbol{\beta_{X_{2}}}$ are correlated is:
\begin{align}
\hat{\theta}_{1ME} &= \frac{\cov_w(\boldsymbol{\hat{\beta}_{Y}}, \boldsymbol{\hat{\beta}_{X_{1}}})\var_w(\boldsymbol{\hat{\beta}_{X_{2}}})-\cov_w(\boldsymbol{\hat{\beta}_{Y}}, \boldsymbol{\hat{\beta}_{X_{2}}})\cov_w(\boldsymbol{\hat{\beta}_{X_{1}}}, \boldsymbol{\hat{\beta}_{X_{2}}})}{\var_w(\boldsymbol{\hat{\beta}_{X_{1}}})\var_w(\boldsymbol{\hat{\beta}_{X_{2}}})-\cov_w(\boldsymbol{\hat{\beta}_{X_{1}}},\boldsymbol{\hat{\beta}_{X_{2}}})^{2}} \notag \\
&\xrightarrow{N \rightarrow \infty} \frac{\cov_w(\boldsymbol{\beta_{Y}}, \boldsymbol{\beta_{X_{1}}})\var_w(\boldsymbol{\beta_{X_{2}}})-\cov_w(\boldsymbol{\beta_{Y}}, \boldsymbol{\beta_{X_{2}}})\cov_w(\boldsymbol{\beta_{X_{1}}}, \boldsymbol{\beta_{X_{2}}})}{\var_w(\boldsymbol{\beta_{X_{1}}})\var_w(\boldsymbol{\beta_{X_{2}}})-\cov_w(\boldsymbol{\beta_{X_{1}}},\boldsymbol{\beta_{X_{2}}})^{2}} \notag \\
&=\theta_1 + \frac{\cov_w(\boldsymbol{\alpha'},\boldsymbol{\beta_{X_{1}}})\var_w(\boldsymbol{\beta_{X_{2}}})-\cov_w(\boldsymbol{\alpha'},\boldsymbol{\beta_{X_{2}}})\cov_w(\boldsymbol{\beta_{X_{1}}},\boldsymbol{\beta_{X_{2}}})}{\var_w(\boldsymbol{\beta_{X_{1}}})\var_w(\boldsymbol{\beta_{X_{2}}})-\cov_w(\boldsymbol{\beta_{X_{1}}},\boldsymbol{\beta_{X_{2}}})^{2}} \label{eq:insidecor}
\end{align}
which is equal to $\theta_1$ if the InSIDE assumption holds with respect to $\boldsymbol{\beta_{X_{1}}}$ and $\boldsymbol{\beta_{X_{2}}}$. As more risk factors with correlated sets of association parameters with $\boldsymbol{\beta_{X_{1}}}$ are included in the multivariable MR-Egger model, additional terms will be added to the bias term in equation~\ref{eq:insidecor}, and the InSIDE assumption must hold for these additional risk factors to obtain a consistent estimate of $\theta_1$.

The variance of the multivariable MR-Egger estimate $\hat{\theta}_{1ME}$ will be heavily influenced by the denominator in the bias term of equation~\ref{eq:insidecor}. As $\boldsymbol{\beta_{X_{1}}}$ and $\boldsymbol{\beta_{X_{2}}}$ become more highly correlated, the standard error of the causal estimate $\hat{\theta}_{1ME}$ will increase, and in some circumstances the estimate from multivariable MR-Egger will be less precise than the estimate from uniavriable MR-Egger. The precision of the causal estimates from multivariable MR-Egger and univariable MR-Egger is discussed further in the Web Appendix. 

\subsection{Advantages of multivariable MR-Egger and comparison with univariable MR-Egger}
\label{sec:advantages}
The bias for the causal estimate from univariable MR-Egger $\hat{\theta}_{UE}$ depends on the weighted covariance between $\boldsymbol{\alpha}$ and $\boldsymbol{\beta_{X_{1}}}$, where:
\begin{equation}
\alpha_j=\alpha'_j+\sum_{i=2}^{K} \theta_{i}\beta_{X_{ij}} \label{eq:pleio}
\end{equation}
The expression in equation~\ref{eq:pleio} follows from the multivariable framework outlined in equation~\ref{eq:multivariable}, where the direct effect for univariable MR-Egger has been decomposed into the residual direct effect $\alpha'_j$ of multivariable MR-Egger and the indirect effects via each risk factor. The residual direct effect $\alpha'_j$ will be altered with each additional risk factor included in the multivariable MR-Egger model. If these additional risk factors are causally associated with the outcome ($\theta_{k}\neq0$), then $\alpha'_j$ will consist of fewer components. It seems likely that the InSIDE assumption would be easier to satisfy for multivariable MR-Egger than its univariable counterpart as the direct effect for univariable MR-Egger consists of unmeasured and measured pleiotropy.

If the $\boldsymbol{\beta_{X_{1}}}$ parameters are independent of the $\boldsymbol{\beta_{X_{k}}}$ parameters for all $k = 2, 3, \ldots, K$, then the second term in equation~\ref{eq:pleio} (the measured direct effect) does not contribute to the value of $\cov_w(\boldsymbol{\alpha}, \boldsymbol{\beta_{X_{1}}})$. Under this scenario, bias for the univariable and multivariable MR-Egger estimates depends on the same covariance term $\cov_w(\boldsymbol{\alpha'}, \boldsymbol{\beta_{X_{1}}})$.  As a consequence, the estimates of the causal effects from univariable MR-Egger $\hat{\theta}_{UE}$ and multivariable MR-Egger $\hat{\theta}_{1ME}$ will be asymptotically the same. In this case, multivariable MR-Egger may improve precision of the causal estimate, but will not affect the asymptotic bias.

When the $\boldsymbol{\beta_{X_{1}}}$ parameters are correlated with at least one of the sets of $\boldsymbol{\beta_{X_{k}}}$ parameters for $k = 2, 3, \ldots, K$, the second term in equation~\ref{eq:pleio} now contributes to the value of $\cov_w(\boldsymbol{\alpha},\boldsymbol{\beta_{X_{1}}})$. The InSIDE assumption for univariable MR-Egger will therefore be automatically violated as the weighted covariance between $\boldsymbol{\alpha}$ and $\boldsymbol{\beta_{X_{1}}}$ will not equal zero, resulting in biased causal estimates of $\theta_1$. If the InSIDE assumption holds for multivariable MR-Egger, and $\boldsymbol{\beta_{X_{k}}}$ are included in the analysis model, then $\hat{\theta}_{1ME}$ will still be a consistent estimate of $\theta_1$. Hence, in this case, multivariable MR-Egger should result in reduced bias compared with univariable MR-Egger.


\subsection{Orientation of the genetic variants}
\label{sec:orientation}
Genetic associations represent the average change in the risk factor or the outcome per additional copy of the reference allele. There is no biological rationale why associations should be expressed with respect to either the major (wildtype) or the minor (variant) allele. In the univariable and multivariable IVW methods, the estimate is not affected by the choice of orientation, as the intercept is fixed at zero. However, in the univariable and multivariable MR-Egger methods, changing the orientation of the variant affects the intercept term and the causal estimate as the orientation affects the definition of the pleiotropy terms $\alpha_j$ and $\alpha'_j$. Consequently, for each choice of orientation, there is a different version of the InSIDE assumption.

To ensure that the MR-Egger analysis does not depend on the reported reference alleles, Bowden \textit{et al.}\ \cite{bowden2015egger} suggested the genetic variants in univariable MR-Egger be orientated so the direction of association with the risk factor is either positive for all variants or negative for all variants. However, this may not be possible for multivariable MR-Egger as the same reference allele must be used for associations with each risk factor and with the outcome. We suggest that the variants should be orientated with respect to their associations with the risk factor of primary interest, although we would recommend a sensitivity analysis considering different orientations if multiple risk factors are of interest. If the genetic variants are all valid instruments, then directional pleiotropy should not be detected with respect to any orientation.
\clearpage
\section{Example: causal effect of HDL-C on CHD risk} 
\label{sec:example}
The effects of high-density lipoprotein cholesterol (HDL-C), low-density lipoprotein cholesterol (LDL-C), and triglycerides on the risk of coronary heart disease (CHD) have been investigated by numerous MR studies \cite{burgess2016lipids}. For HDL-C, univariable MR suggested a causally protective role against CHD risk, whereas univariable MR-Egger provided no evidence of a causal effect and the test for directional pleiotropy was statistically significant at the 5\% level \cite{bowden2016consistent}. A null causal effect for HDL-C was also reported from a multivariable MR analysis that included LDL-C and triglycerides using the multivariable IVW method \cite{burgess2015multivariable}, although a small but protective causal effect was estimated in a further multivariable MR analysis using a wider range of 185 genetic variants \cite{burgess2014multilipid}.

We investigate the causal effect of HDL-C on CHD risk further using the multivariable MR-Egger method.  We consider the 185 genetic variants having known association with at least one of HDL-C, LDL-C and triglycerides at GWAS significance in 188,578 participants reported by the Global Lipids Genetics Consortium \cite{willer2013}. The point estimates for the associations between these genetic variants and lipids were taken from Do \textit{et al.}\ \cite{do2013}. The CARDIoGRAMplusC4D consortium consisting of 60,801 cases and 123,504 controls was used to obtain the estimates of the association between the variants and CHD risk \cite{nikpay2015}. The IVW and MR-Egger methods were applied to the data under univariable and multivariable frameworks as described in Section~\ref{sec:methods}. For the univariable IVW and MR-Egger methods, the models were fitted using two sets of variants: firstly using all 185 variants; and secondly using all variants associated with HDL-C at GWAS level of significance. The genetic variants were orientated with respect to the risk increasing allele for HDL-C. These analyses differ from those provided in \cite{burgess2014multilipid} and \cite{do2013} as they use summarized data from different versions of the CARDIoGRAMplusC4D study; here we use associations from the 2015 data release \cite{nikpay2015}.

The univariable IVW method suggested a significant protective effect of HDL-C for both sets of variants with a causal odds ratio of 0.88 (95\% CI: 0.80, 0.97) for all variants (Table~\ref{tab:hdl}). This estimate attenuated to the null in the univariable MR-Egger method (0.98, 95\% CI: 0.87, 1.11) with evidence of directional pleiotropy (p-value=0.004). The causal odds ratios from multivariable IVW (0.96, 95\% CI: 0.89, 1.05) and multivariable MR-Egger (1.04, 95\% CI: 0.94, 1.14) had opposite directions of association, with both analyses indicating that HDL-C is not causally associated with CHD risk. The significant result for directional pleiotropy in the multivariable MR-Egger method suggests that LDL-C and triglycerides do not fully explain the direct effects of the genetic variants on the outcome, suggesting that there is still residual pleiotropy via other unmeasured risk factors.  

\begin{center}
[Table 1 should appear about here.]
\end{center}

\subsection{Varying the orientation of the genetic variants}
As a sensitivity analysis, the multivariable MR-Egger method was re-performed with the genetic variants orientated with respect to the risk increasing alleles for LDL-C and triglycerides.  

The causal estimates for HDL-C, LDL-C, and triglycerides from multivariable MR-Egger when the variants were orientated with respect to HDL-C, LDL-C or triglycerides are presented in Table~\ref{tab:orientation}. Estimates of the MR-Egger intercept are also provided for the three models. To allow for comparisons between the multivariable methods, the causal estimates from multivariable IVW are included in Table~\ref{tab:orientation}. The causal estimates in bold follow the recommendation outlined in Section~\ref{sec:orientation} that the genetic variants should be orientated with respect to the risk factor-increasing allele for the risk factor of interest.  

All of the causal odds ratios for HDL-C from the multivariable MR-Egger models indicated that HDL-C is not causally associated with CHD risk.  
Significant adverse effects of LDL-C on CHD risk were reported from the multivariable IVW (1.45, 95\% CI: 1.34, 1.58) and multivariable MR-Egger (1.52, 95\% CI: 1.37, 1.69) methods. Orientating the variants with respect to the risk increasing alleles for HDL-C and triglycerides had little impact on the causal estimates for LDL-C from multivariable MR-Egger. The multivariable IVW method suggested a significant adverse effect of triglycerides on CHD risk with a causal odds ratio of 1.19 (95\% CI: 1.07, 1.33), this estimate was attenuated to the null in the multivariable MR-Egger method (1.09, 95\% CI: 0.96, 1.23). The causal odds ratios for triglycerides remained significant, however, when the variants were orientated with respect to HDL-C and LDL-C in the multivariable MR-Egger models.   

Since the orientation of the genetic variants affects the interpretation of the direct effect, and the definition of the InSIDE assumption, the MR-Egger intercept will vary between different orientations. In this example, the MR-Egger intercept differed from zero when the variants were orientated with respect to HDL-C and triglycerides, yet there was no evidence of directional pleiotropy or the InSIDE assumption being violated when the variants were orientated with respect to LDL-C.  

\begin{center}
[Table 2 should appear about here.]
\end{center}

\pagebreak
\section{Simulation study}
\label{sec:simulation}
In order to assess the merits of using multivariable MR-Egger over multivariable IVW and univariable MR-Egger in realistic settings, we perform a simulation study. Univariable and multivariable MR-Egger will be compared with respect to the consistency of the causal estimates and statistical power to detect the causal effect.  The setup of the simulation study corresponds to the applied example in Section~\ref{sec:example} and will be considered under two broad scenarios: (1) $\boldsymbol{\beta_{X_{k}}}$ are generated independently for all $k = 1, 2, \ldots, K$; and (2) $\boldsymbol{\beta_{X_{k}}}$ are correlated for all $k = 1, 2, \ldots, K$.  Estimates of the $R^{2}$ and F-statistic for the applied example are provided in the Web Appendix.\\

We simulated summarized level data for 185 genetic variants indexed by $j=1,2, \ldots, J$ for three risk factors ($X_{1}$, $X_{2}$, $X_{3}$) and an outcome $Y$ from the following data-generating model:
\begin{align}
\left(\begin{matrix}
\beta_{X_{1j}} \\
\beta_{X_{2j}} \\
\beta_{X_{3j}}
\end{matrix}\right)
&\sim \mathcal{N}_3\left(\left(\begin{matrix}
0.08 \\
0.03 \\
-0.05 \\
\end{matrix}\right),\left(\begin{matrix}
{\sigma_{1}}^{2} & \rho_{12}\sigma_{1}\sigma_{2} & \rho_{13}\sigma_{1}\sigma_{3} \\
\rho_{12}\sigma_{1}\sigma_{2} & {\sigma_{2}}^{2} & \rho_{23}\sigma_{2}\sigma_{3} \\
\rho_{13}\sigma_{1}\sigma_{3} & \rho_{23}\sigma_{2}\sigma_{3} & {\sigma_{3}}^{2}\\
\end{matrix}\right) \right) \notag \\
\beta_{Yj} &= \alpha'_{j} + \theta_{1}|\beta_{X_{1j}}| + \theta_{2}\beta_{X_{2j}} + \theta_{3}\beta_{X_{3j}} +\epsilon_{j} \notag \\
\epsilon_{j} &\sim \mathcal{N}(0,1) \notag \\
\alpha'_{j} &\sim \mathcal{N}(\mu,0.004) \label{eq:datgen}
\end{align}
The primary objective was to estimate $\theta_{1}$, with the causal effects set to: $\theta_{1}=0$ (null causal effect) or $\theta_{1}=0.3$ (positive causal effect); $\theta_{2}=0.1$; and $\theta_{3}=-0.3$. The data were simulated to consider the following four scenarios: \\

\begin{compactitem}
\item[1.] No pleiotropy ($\alpha'_{j}=0$ for all $j$), InSIDE assumption automatically satisfied;
\item[2.] Balanced pleiotropy ($\mu=0$), InSIDE assumption satisfied;
\item[3.] Directional pleiotropy ($\mu=0.01, 0.05$ or $0.1$), InSIDE assumption satisfied;
\item[4.] Directional pleiotropy ($\mu=0.01, 0.05$ or $0.1$), InSIDE assumption violated. \\
\end{compactitem}
When the InSIDE assumption for multivariable MR-Egger was satisfied, $\alpha'_{j}$ and $\beta_{X_{1j}}$ were drawn from independent distributions, and when it was violated they were drawn from a multivariate normal distribution with $\cor(\boldsymbol{\alpha'},\boldsymbol{\beta_{X_{1}}})=0.3$. The above four scenarios were applied to the simulated data when $\boldsymbol{\beta_{X_{k}}}$ were generated independently for all $k$, with the parameters in the covariance matrix set to: $\sigma_{1}^{2}=0.03$; $\sigma_{2}^{2}=0.02$; $\sigma_{3}^{2}=0.04$; and $\rho_{12}=\rho_{13}=\rho_{23}=0$. The four scenarios were repeated when $\boldsymbol{\beta_{X_{k}}}$ were correlated for all $k$ ($\rho_{12}=0.2$, $\rho_{13}=-0.3$, $\rho_{23}=0.1$). In total, data were simulated for 32 different choices of parameters.  

To ensure the direction of association between $G_{j}$ and $X_{1}$ was the same for all $j$ variants, the absolute value of the genetic associations with $X_{1}$ ($|\beta_{X_{1j}}|$) were used to generate $\beta_{Y_{j}}$ (equation \ref{eq:datgen}). It was assumed that $\beta_{X_{kj}}$ (for all $k$) and $\beta_{Y_{j}}$ had the same reference allele and the genetic variants were uncorrelated. The multivariable IVW, univariable MR-Egger and multivariable MR-Egger methods were applied to the simulated datasets.  The weights for the multivariable IVW and multivariable MR-Egger are given by equation \ref{wts: MV}, while equation \ref{wts: UV} contains the weights for univariable MR-Egger.
\begin{align}
\se(\beta_{Y_{j}})^{-2}&={({\epsilon_{j}}^{2}+{\sigma_{\alpha'}}^{2})}^{-1} \label{wts: MV} \\
\se(\beta_{Y_{j}})^{-2}&={({\epsilon_{j}}^{2}+{\sigma_{\alpha'}}^{2}+{\theta_{2}}^{2}{\sigma_{2}}^{2} + {\theta_{3}}^{2}{\sigma_{3}}^{2})}^{-1} \label{wts: UV}
\end{align}

\subsection{Results} 
\label{sec:simulation_results}
\label{sec:results}
The results from the simulation study using 10\thinspace000 simulated datasets are presented in Table~\ref{tab:indep} ($\boldsymbol{\beta_{X_{k}}}$ generated independently) and Table \ref{tab:cor} ($\boldsymbol{\beta_{X_{k}}}$ correlated). For each scenario, the mean estimate, the mean standard error, and the statistical power to detect a null or positive causal effect at a nominal 5\% significance level are presented in Tables~\ref{tab:indep} and \ref{tab:cor} for the multivariable IVW, univariable MR-Egger and multivariable MR-Egger methods. For univariable and multivariable MR-Egger, the statistical power of the MR-Egger intercept test is also provided.  

\textbf{$\boldsymbol{\beta_{X_{k}}}$ generated independently:} In scenarios 1 and 2 (no and balanced pleiotropy), estimates from all methods were unbiased, and those from the multivariable IVW method were the most precise. In scenarios 3 and 4 (directional pleiotropy), estimates from the multivariable IVW method were biased, with the magnitude of bias increasing as the average value of $\boldsymbol{\alpha'}$ increased from 0.01 to 0.1. In scenario 3 (InSIDE satisfied), estimates from the univariable and multivariable MR-Egger methods were unbiased, whereas in scenario 4 (InSIDE violated), they were biased. Although the causal estimates for both multivariable IVW and multivariable MR-Egger were biased under scenario 4, the magnitude of bias was less for multivariable MR-Egger, with the exception of when $\alpha'_j$ was generated from $\mathcal{N}(0.01,0.004)$. Precision and power to detect a causal effect were always better for the multivariable MR-Egger method than univariable MR-Egger, although the univariable MR-Egger method detected directional pleiotropy more often. The average value of $\boldsymbol{\alpha'}$ had no impact on the degree of bias for univariable or multivariable MR-Egger.

\textbf{$\boldsymbol{\beta_{X_{k}}}$ correlated:} Bias for the multivariable IVW method was present in scenarios 3 and 4 only, as in the independently generated setting. In this setting, the InSIDE assumption for univariable MR-Egger was violated for all four scenarios, resulting in biased point estimates of $\theta_1$. However, the multivariable InSIDE assumption was satisfied for scenarios 1, 2 and 3, and so causal estimates from multivariable MR-Egger were unbiased. When the multivariable InSIDE assumption was violated (scenario 4) the estimates from multivariable MR-Egger were biased, yet the magnitude of bias was less compared with univariable MR-Egger as $|\cov(\boldsymbol{\alpha'},\boldsymbol{\beta_{X_{1}}})|< |\cov(\boldsymbol{\alpha},\boldsymbol{\beta_{X_{1}}})|$.

\begin{center}
[Table 3 should appear about here.]
\end{center}

\begin{center}
[Table 4 should appear about here.]
\end{center}

\subsection{Causal relationships between the risk factors} 
The simulations performed in Section~\ref{sec:simulation_results} assumed that the effect of each risk factor on the outcome is not mediated through another risk factor. There may be circumstances where causal relationships between risk factors are biologically plausible. Burgess \textit{et al.}\ \cite{burgess2015multivariable} illustrated that the multivariable IVW method estimates the direct causal effects ($\theta_k$) of each risk factor on the outcome, irrespective of whether causal relationships between the risk factors exist. 

In the applied example of the paper, there may also be deterministic dependencies between the risk factors. LDL-C is rarely measured directly, but is estimated from measurements of total cholesterol, triglycerides and HDL-C via the Friedewald equation as total cholesterol minus HDL-C minus 0.2 times triglycerides (assuming all measurements in mg/dL) \cite{Friedewald1972}. It has previously been shown that the coefficient for LDL-C is the same as the coefficient for non-HDL-C (calculated as total cholesterol minus HDL-C) in a regression model including HDL-C and triglycerides (see Appendix 2 in \cite{Angelantonio2009}). However, the coefficient for triglycerides will change, as the non-HDL-c measure contains more triglycerides than the LDL-c measure. Hence, in the case that there are deterministic relationships between the risk factors, effect estimates may change as the choice of risk factors varies due to their interpretation as direct effects conditional on other risk factors in the regression model.

\begin{figure}[htb]
\centering
\includegraphics[scale=0.15]{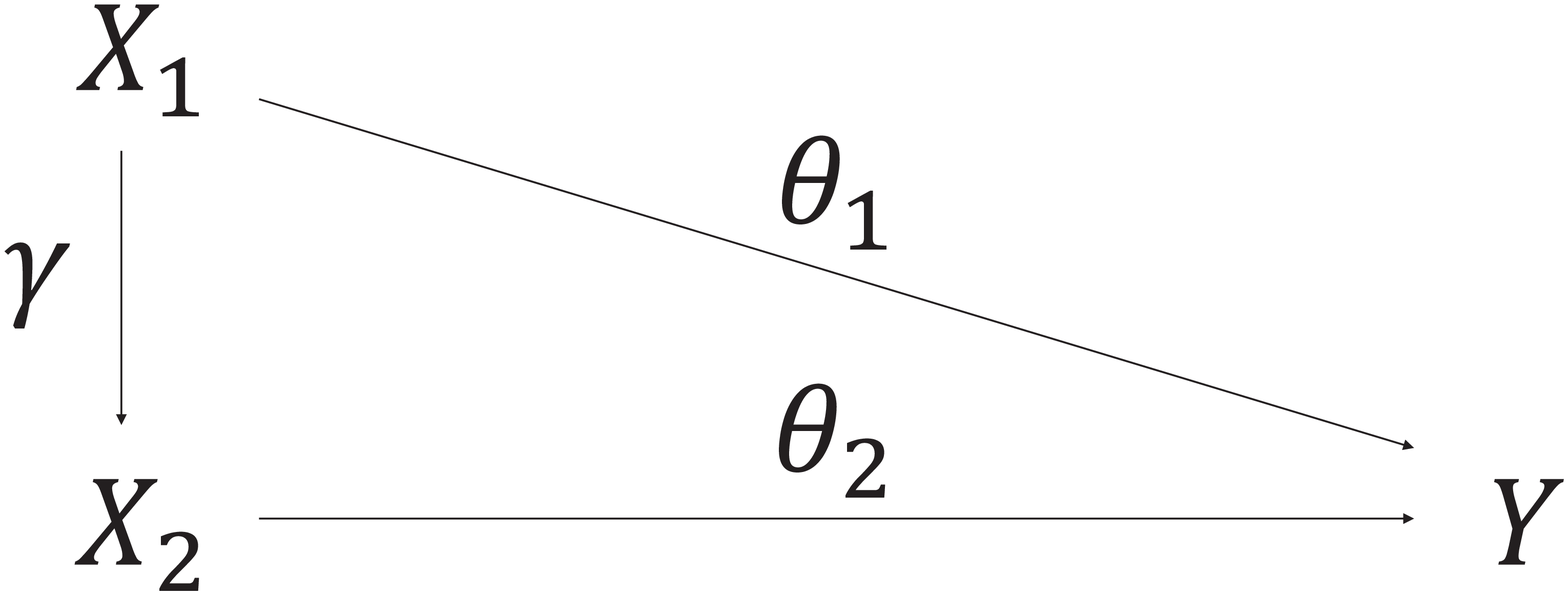}
\caption{Causal directed acyclic graph illustrating the causal relationships between the two risk factors $X_{1}$ and $X_{2}$, and outcome $Y$. The causal effect of $X_{1}$ on $X_{2}$ is $\gamma$, and the direct causal effect of the risk factor $X_k$ on the outcome $Y$ is $\theta_{k}$. The total causal effect of $X_1$ on $Y$ is $\theta_1 + \gamma\theta_2$; consisting of the direct effect ($\theta_1$) and the indirect effect via $X_2$ ($\gamma\theta_2$). $U_k$ represents the set of variables that confound the associations between $X_k$ and $Y$.} \label{dag3}
\end{figure}

We performed additional simualtions to investigate the behaviour of the multivariable MR-Egger method when $X_2$ is causally dependent on $X_1$, and the causal effect of $X_1$ on $X_2$ is $\gamma$ (Figure~\ref{dag3}). The total causal effect of $X_1$ on $Y$ is $\theta_1 + \gamma\theta_2$; consisting of the direct effect ($\theta_1$) and the indirect effect via $X_2$ ($\gamma\theta_2$). See the Web Appendix for more details on the data generating model.  

\begin{center}
[Figure 3 should appear about here.]
\end{center}

\subsubsection{Results}
The results from the additional simulations are provided in Web Table~\ref{tab:indep_2} and Web Table~\ref{tab:cor_2}. In scenarios where there was no bias in the original set of simulations, the multivariable IVW and multivariable MR-Egger methods consistently estimated the direct effect of $X_1$ on $Y$ ($\theta_1$), whilst the univariable MR-Egger method consistently estimated the total causal effect of $X_1$ on $Y$ ($\theta_1 + \gamma\theta_2$). Compared to the results in Section~\ref{sec:simulation_results}, precision and power to detect a causal effect were reduced for the multivariable IVW and multivariable MR-Egger methods. This reduction in power was anticipated since the multivariable models condition on the mediator along a causal pathway, which is known to decrease power to detect a causal effect \cite{Fritz2007}.  

\pagebreak
\section{Discussion}
\label{sec:discussion}
In this paper we have extended univariable MR-Egger to the multivariable setting and outlined the assumptions required to obtain consistent causal estimates in the presence of directional pleiotropy. Multivariable MR-Egger should be viewed as a sensitivity analysis to provide robustness against both measured and unmeasured pleiotropy, and to strengthen the evidence from the original MR analysis. If the causal estimate from multivariable MR-Egger is substantially different from the estimate obtained in the original analysis, then further investigation into the causal finding and the potential for pleiotropy is required.

The simulation study has highlighted the benefits of using multivariable MR-Egger over its univariable counterpart. This is particularly true when the associations of the genetic variants with the risk factor of interest are associated with genetic associations with at least one of the risk factors (measured pleiotropy). Under this scenario, the InSIDE assumption for univariable MR-Egger is likely to be violated, leading to biased causal estimates. Multivariable MR-Egger will, however, produce consistent causal estimates if the InSIDE assumption for multivariable MR-Egger is satisfied.  Although the estimates from univariable and multivariable MR-Egger are asymptotically the same when genetic associations with each risk factor are all independent, multivariable MR-Egger should also have greater power to detect a causal effect when the InSIDE assumption is satisfied. Given these advantages, and the sensitivity of the multivariable IVW method to directional pleiotropy, we believe that multivariable MR-Egger should be considered as an important sensitivity analysis for a MR study.

\subsection{Multivariable by design, or multivariable as a sensitivity analysis?}
There are two possible scenarios where multivariable MR-Egger may be used as a sensitivity analysis: either the primary analysis is considered to be multivariable by design, or a multivariable framework is only considered as part of the sensitivity analysis. The first case should be motivated by biological evidence where the set of risk factors are known to be associated with common genetic variants, such as lipid fractions. Under this scenario, multivariable IVW should be used as the primary analysis method with multivariable MR-Egger providing robustness against directional pleiotropy as a sensitivity analysis.

In the second scenario, where there is a lack of biological evidence to suggest a multivariable framework, univariable IVW would generally be considered as the primary analysis method and univariable MR-Egger as the main sensitivity analysis. However, if the genetic variants are associated with other risk factors, multivariable MR-Egger could also be used as a sensitivity analysis as its assumptions are more likely to be satisfied and it may have greater power to detect a causal effect than univariable MR-Egger. An example of the use of multivariable MR as a sensitivity analysis is an MR study on plasma urate concentrations and CHD risk \cite{white2016urate}. To account for measured and unmeasured pleiotropic associations of the genetic variants, the authors performed the multivariable IVW and univariable MR-Egger methods as sensitivity analyses. This investigation may have benefited from performing the multivariable MR-Egger method to simultaneously account for both measured and unmeasured pleiotropic associations.

\subsection{InSIDE assumption and orientation of genetic variants}
The validity of multivariable MR-Egger and its ability to estimate consistent causal effects is dependent upon the InSIDE assumption being satisfied. Whilst it is not possible to determine whether the InSIDE assumption has been violated, we believe it is more likely to hold for multivariable MR-Egger then univariable MR-Egger. When the $\boldsymbol{\beta_{X_{1}}}$ parameters are associated with at least one of the sets of $\boldsymbol{\beta_{X_{k}}}$  parameters for $k = 2, 3, \ldots, K$, the InSIDE assumption for univariable MR-Egger is automatically violated and causal estimates from the method will be inconsistent. The direct effects of the genetic variants on the outcome will consist of fewer components for multivariable MR-Egger compared to its univariable counterpart, making it more plausible that the InSIDE assumption will hold for multivariable MR-Egger.

The recommendation of orientating the genetic variants in multivariable MR-Egger to the risk factor-increasing or risk factor-decreasing allele for the risk factor of interest may be considered arbitrary. While we accept this limitation, we would argue it brings consistency to the results. This recommendation may result in the analysis being performed up to $K$ times to obtain the causal estimates for all $K$ risk factors. The orientation of the genetic variants will also affect the interpretation of the direct effect, thereby altering the InSIDE assumption. This may result in the MR-Egger intercept estimate varying between different orientations. This was seen in the applied example where the intercept term was non-significant when the alleles were orientated with respect to LDL-C, and significant when orientated with respect to HDL-C and trigclyercides.

\subsection{Linearity and homogeneity assumptions}
Throughout this paper we have assumed linearity and homogeneity (no effect modification) of the causal effects of the risk factors on the outcome, and of the associations between the genetic variants with the risk factors and with the outcome. If the assumptions of linearity and homogeneity are violated then the methods discussed in this paper still provide a valid test for the null hypothesis of whether the risk factor is causally associated with the outcome \cite{burgess2016multiple}. The causal estimate, however, would not have a literal interpretation if the assumptions were violated \cite{burgess2015beyond}. Although linearity and homogeneity are strong assumptions, the effect of genetic variants on the risk factor and outcome tend to be limited to a small range, which may make the assumptions of linearity and homogeneity more reasonable in an MR analysis.

The multivariable models have assumed that the risk factors do not have causal effects on each other. The additional simulation study has illustrated that the multivariable MR-Egger method estimates the direct causal effects of the risk factors on the outcome, irrespective of whether the risk factors are causally related. There was, however, a reduction in precision and power to detect the causal effect for multivariable MR-Egger when a causal relationship between the risk factors was present. Conversely, univariable MR-Egger will produce consistent causal estimates of the total effect if the InSIDE assumption for univariable MR-Egger is satisfied.  

\subsection{Implication for future research}
The paper by Helgadottir \textit{et al.}\ \cite{helgadottir2016} highlights the importance and need to develop sensitivity analyses for multivariable MR. This is particularly relevant given the recent advances in high-throughput phenotyping which has led to the introduction of `-omics' data such as metabolomics, genomics, and proteomics \cite{relton2012epigenetics}. Genome-wide analyses of high-dimensional `-omics' data are becoming more popular \cite{suhre2016,kastenmuller2015}, yet few MR analyses have been performed using these datasets \cite{burgess2016lipids}. As summarized data from large consortia become more accessible, the opportunities to use MR on high-dimensional datasets will only increase.  Methods such as multivariable MR-Egger will be valuable to investigate the causal effects of multiple related phenotypes with shared genetic predictors.  

Bowden \textit{et al} \cite{bowden2016I_2} have shown that uncertainty in the associations between the genetic variants and the risk factor in univariable MR-Egger can lead to attenuation towards the null for a positive causal effect.  This attenuation is approximately equal to the $I^2$ statistic from meta-analysis of the weighted associations with the exposure $\hat{\beta}_{Xj} \se(\hat{\beta}_{Yj})^{-1}$ with standard errors $\se(\hat{\beta}_{Xj}) \se(\hat{\beta}_{Yj})^{-1}$ \cite{bowden2016I_2}. It is unclear whether the method proposed by Bowden \textit{et al} \cite{bowden2016I_2} can be directly applied to the multivariable setting. The effect of regression dilution bias in a multivariable regression model can lead to the association being over-estimated, under-estimated or the direction of association being reversed \cite{Frost2005}. However, in the simulation study considered in this paper,
substantial bias due to uncertainty in the genetic associations with the
risk factor was not observed. Further research is required to investigate this issue. 

Throughout the paper, we have assumed that the genetic variants are uncorrelated (not in linkage disequilibrium). This assumption, and the requirement for further methodological development, is discussed in the Web Appendix. 

\subsection*{Acknowledgements}
Jessica Rees is supported by the British Heart Foundation (grant number FS/14/59/31282). Stephen Burgess is supported by Sir Henry Dale Fellowship jointly funded by the Wellcome Trust and the Royal Society (Grant Number 204623/Z/16/Z). 

\clearpage
\putbib[ref_reduced]
\end{bibunit}

\clearpage
\subsection*{Tables}
\begin{table}[h] 
\begin{center}
\begin{small}
\centering
\caption{Log causal odds ratios (95\% confidence intervals) for coronary heart disease per standard deviation increase in HDL-C, with two-sided p-values. Estimates of the intercept are given in univariable and multivariable MR-Egger.}
\resizebox{\textwidth}{!} { 
\begin{tabular}[c]{lR*{7}{c}} 
\toprule
\multicolumn{2}{c}{} &
\multicolumn{3}{c}{\textbf{Causal estimate}} &
\multicolumn{3}{c}{\textbf{MR-Egger intercept test}} \\
\cmidrule(r){3-5}\cmidrule(r){6-8}
& & $\hat{\theta}_{\mbox{HDL-C}}$ (CI) & $\se(\hat{\theta}_{\mbox{HDL-C}})$ & p-value &       $\hat{\theta}_{\mbox{0E}}$ & $\se(\hat{\theta}_{\mbox{0E}})$ & p-value \\ [0.1cm]
\toprule
\multicolumn{8}{l}{\textbf{Univariable IVW}} \\
All variants  & & -0.130 (-0.227, -0.033) &	0.049 & 0.009 & - & - & -\\
Reduced set of variants$^{\mbox{a}}$ & & -0.114 (-0.211, -0.017) & 0.049 & 0.022 & - & - & -\\
\rule{0pt}{1ex} \\

\multicolumn{8}{l}{\textbf{Univariable MR-Egger}} \\
All variants & & -0.016 (-0.138, 0.106) & 0.062 & 0.800	& -0.007 &	0.002 & 0.004  \\
Reduced set of variants$^{\mbox{a}}$ & & 0.067 (-0.070, 0.204) &	0.069 & 0.332 &	-0.012 & 0.004	& 0.001 \\
\rule{0pt}{1ex} \\

\textbf{Multivariable IVW} & & -0.039 (-0.123, 0.045) & 0.042	& 0.359	& -	& -	& - \\
\rule{0pt}{1ex} \\

\textbf{Multivariable MR-Egger} & & 0.036 (-0.063, 0.134) & 0.050 & 0.477 &	-0.005 & 0.002 & 0.008 \\
\bottomrule
\end{tabular}
}
\caption*{Abbreviations: MR, Mendelian randomization; HDL-C, high-density lipoprotein cholesterol; IVW, inverse-variance weighted; CI, confidence interval; SE, standard error.\\
$^{\mbox{a}}$95 variants associated with HDL-C at a genome-wide level of significance (p-value$<5\times10^{-8}$).} \label{tab:hdl}
\end{small} %
\end{center}
\end{table} 

\begin{table}[h] 
\begin{center}
\begin{small}
\centering
\caption{Causal log odds ratios (95\% confidence intervals) for coronary heart disease per standard deviation increase in HDL-C, LDL-C, and triglycerides from multivariable IVW and multivariable MR-Egger. Estimates from multivariable MR-Egger are presented from three models where the reference allele is the risk increasing allele for HDL-C, LDL-C or triglycerides. Estimates of the intercept are given for multivariable MR-Egger.}
\resizebox{\textwidth}{!} { 
\begin{tabular}[c]{lR*{5}{c}} 
\toprule
\multicolumn{2}{c}{} &
\multicolumn{3}{c}{\textbf{Causal estimates}} &
\multicolumn{1}{c}{\textbf{MR-Egger intercept}} \\
\cmidrule(r){3-5} \cmidrule(r){6-6}
\multicolumn{2}{c}{} &
\multicolumn{1}{c}{\boldmath$\hat{\theta}_{\mbox{HDL-C}}$} &
\multicolumn{1}{c}{\boldmath$\hat{\theta}_{\mbox{LDL-C}}$} &
\multicolumn{1}{c}{\boldmath$\hat{\theta}_{\mbox{TG}}$} &
\multicolumn{1}{c}{\boldmath$\hat{\theta}_{0E}$} \\ [0.1cm]
\toprule
\textbf{Multivariable IVW} & & -0.039 (-0.123, 0.045) & 0.375 (0.292,	0.457) & 0.173 (0.063, 0.283) & - \\
\rule{0pt}{1ex} \\
\textbf{Multivariable MR-Egger}\\
Orientation with respect to$^{\mbox{a}}$:\\
HDL-C & & \textbf{0.036 (-0.063, 0.134)} & 0.378 (0.297, 0.458) & 0.136 (0.024, 0.247) & -0.005 (-0.009, -0.001) \\
LDL-C & & -0.034 (-0.118, 0.049) & \textbf{0.420 (0.318, 0.522)} &	0.194 (0.081, 0.308) & -0.003 (-0.007, 0.001)\\
TG & & -0.018 (-0.102, 0.066) & 0.350 (0.267, 0.433) & \textbf{0.083 (-0.045,	0.211)} & 0.005 (0.001, 0.009)\\
\bottomrule
\end{tabular}
}
\caption*{Abbreviations: MR, Mendelian randomization; HDL-C, high-density lipoprotein cholesterol; LDL-C, low-density lipoprotein cholesterol; TG, triglycerides.\\
$^{\mbox{a}}$Alleles orientated for all genetic associations with respect to the risk increasing allele for HDL-C, LDL-C or triglycerides.} \label{tab:orientation}
\end{small} %
\end{center}
\end{table}

\pagebreak
\begin{table}[htbp] 
\begin{center}
\begin{small}
\centering
\caption{Performance of multivariable IVW, univariable MR-Egger and multivariable MR-Egger with respect to $\hat{\theta}_{1}$ for a null ($\theta_{1}=0$) and positive ($\theta_{1}=0.3$) causal effect where $\boldsymbol{\beta_{X_{k}}}$ are generated independently for all $k$. All tests were performed at the 5\% level of significance.}
\resizebox{\textwidth}{!} {
\begin{tabular}[c]{lR*{8}{c}}
\toprule
\multicolumn{2}{c}{} &
\multicolumn{2}{c}{\textbf{Multivariable IVW}} &
\multicolumn{3}{c}{\textbf{Univariable MR-Egger}} &
\multicolumn{3}{c}{\textbf{Multivariable MR-Egger}} \\
\cmidrule(r){3-4}\cmidrule(r){5-7}\cmidrule(r){8-10}
& & Mean $\hat{\theta}_{1}$ & Power, & Mean $\hat{\theta}_{1}$ & \multicolumn{2}{c}{Power, \%} & Mean $\hat{\theta}_{1}$ & \multicolumn{2}{c}{Power, \%} \\
& & (mean SE) & \% & (mean SE) & Intercept & Causal &
(mean SE) & Intercept & Causal \\
\toprule
\multicolumn{9}{l}{\textbf{Null causal effect: \boldmath$\theta_{1}=0$}} \\
\multicolumn{9}{l}{\underline{1. No pleiotropy, InSIDE satisfied}} \\
 				&	& 0.000 (0.045) 	& 3.8 	& -0.002 (0.158) & 9.1 & 4.7 & 0.000 (0.084) 	& 3.7 	& 4.1 \\
\rule{0pt}{1ex} \\ 	
							
\multicolumn{9}{l}{\underline{2. Balanced pleiotropy, InSIDE satisfied}} \\
\plei{}(0,0.004) &	& -0.001 (0.100) 	& 4.7 	& -0.001 (0.187) & 7.8 & 4.7 & 0.000 (0.165) 	& 4.6 	& 4.6 \\
\rule{0pt}{1ex} \\

\multicolumn{9}{l}{\underline{3. Directional pleiotropy, InSIDE satisfied}} \\
\plei{}(0.01,0.004) &	& 0.041 (0.100) 	& 6.7 	& -0.003 (0.187)	& 12.2	& 4.3	& -0.002 (0.165)	& 5.9	& 4.5 \\
\plei{}(0.05,0.004)&	& 0.210 (0.100)		& 55.3	& 0.002 (0.187)	  & 49.2	& 4.6	& 0.002 (0.166)	& 36.3	& 4.6 \\
\plei{}(0.1,0.004) &	& 0.417 (0.102) 	& 97.4	& 0.000 (0.187)	
& 91.6	& 4.3	& 0.001 (0.165)	& 88.0	& 4.6 \\
\rule{0pt}{1ex} \\

\multicolumn{9}{l}{\underline{4. Directional pleiotropy, InSIDE violated}} \\
\plei{}(0.01,0.004) &	& 0.074 (0.100) 	& 12.3	& 0.089 (0.187)
& 6.7 	& 7.6	& 0.088 (0.165)	& 4.3	& 8.4 \\
\plei{}(0.05,0.004) &	& 0.240 (0.100)		& 67.2	& 0.089 (0.187)	
& 34.1 	& 7.8	& 0.088 (0.165)	& 21.1	& 8.8 \\
\plei{}(0.1,0.004) 	&	& 0.450 (0.101)		&98.6	& 0.088 (0.187)
& 84.1 	& 7.6	& 0.088 (0.165)	& 78.7	& 8.7 \\
\rule{0pt}{1ex} \\

\multicolumn{9}{l}{\textbf{Positive causal effect: \boldmath$\theta_{1}=0.3$}} \\
\multicolumn{9}{l}{\underline{1. No pleiotropy, InSIDE satisfied}} \\
				&	& 0.300 (0.044)		& 98.9	& 0.300 (0.157) & 9.3
& 50.1	& 0.300 (0.084)	& 4.3	& 87.3 \\
\rule{0pt}{1ex} \\

\multicolumn{9}{l}{\underline{2. Balanced pleiotropy, InSIDE satisfied}} \\
\plei{}(0,0.004) &	& 0.301 (0.100)		& 84.6	& 0.303 (0.187) & 7.5
& 38.2	& 0.302 (0.166)	& 4.9	& 46.4 \\
\rule{0pt}{1ex} \\

\multicolumn{9}{l}{\underline{3. Directional pleiotropy, InSIDE satisfied}} \\
\plei{}(0.01,0.004) &	& 0.343 (0.100)		& 91.5	& 0.300 (0.187)	
& 12.8	& 36.8	& 0.299 (0.165) & 6.0	& 45.8 \\
\plei{}(0.05,0.004) &	& 0.509 (0.100)		& 99.7	& 0.300	(0.188)	
& 50.6	& 37.3	& 0.299 (0.166)	& 37.1	& 46.1 \\
\plei{}(0.1,0.004) 	&	& 0.716 (0.102)		& 100.0	& 0.300 (0.187)	
& 91.1	& 37.1	& 0.299 (0.166)	& 87.9	& 46.1 \\
\rule{0pt}{1ex} \\

\multicolumn{9}{l}{\underline{4. Directional pleiotropy, InSIDE violated}} \\
\plei{}(0.01,0.004) &	& 0.374 (0.099)		& 94.3	& 0.390 (0.187)	
& 6.6	& 56.4	& 0.389 (0.165)	& 4.6	& 65.8 \\
\plei{}(0.05,0.004) &	 & 0.539 (0.100)	& 99.8	& 0.388 (0.187)	
& 34.4	& 55.6	& 0.387 (0.165)	& 21.5	& 65.5 \\
\plei{}(0.1,0.004) &	& 0.747 (0.101)		& 100.0	& 0.383 (0.187)	
& 84.7	& 55.1	& 0.384 (0.165)	& 78.3	& 65.2 \\
\bottomrule
\end{tabular}
}
\caption*{Abbreviations: MR, Mendelian randomization; SE, standard error; IVW, inverse-variance weighted; InSIDE, Instrument Strength Independent of Direct Effect.} \label{tab:indep}
\end{small} %
\end{center}
\end{table}

\pagebreak
\begin{table}[h] 
\begin{center}
\begin{small}
\centering
\caption{Performance of multivariable IVW, univariable MR-Egger and multivariable MR-Egger with $\boldsymbol{\beta_{X_{k}}}$ being correlated for all $k$.}
\resizebox{\textwidth}{!} {
\begin{tabular}[c]{lR*{8}{c}}
\toprule
\multicolumn{2}{c}{} &
\multicolumn{2}{c}{\textbf{Multivariable IVW}} &
\multicolumn{3}{c}{\textbf{Univariable MR-Egger}} &
\multicolumn{3}{c}{\textbf{Multivariable MR-Egger}} \\
\cmidrule(r){3-4}\cmidrule(r){5-7}\cmidrule(r){8-10}
& & Mean $\hat{\theta}_{1}$ & Power, & Mean $\hat{\theta}_{1}$ & \multicolumn{2}{c}{Power, \%} & Mean $\hat{\theta}_{1}$ & \multicolumn{2}{c}{Power, \%} \\
& & (mean SE) & \% & (mean SE) & Intercept & Causal &
(mean SE) & Intercept & Causal \\
\toprule
\multicolumn{9}{l}{\textbf{Null causal effect: \boldmath$\theta_{1}=0$}} \\
\multicolumn{9}{l}{\underline{1. No pleiotropy, InSIDE satisfied}} \\
 				&	& 0.000 (0.047) 	& 4.0 	& 0.099 (0.157) & 4.3 & 10.1 & 0.000 (0.086) 	& 4.4 	& 4.6 \\
\rule{0pt}{1ex} \\ 	
							
\multicolumn{9}{l}{\underline{2. Balanced pleiotropy, InSIDE satisfied}} \\
\plei{}(0,0.004) &	& -0.001 (0.104) 	& 4.7 	& 0.093 (0.187) & 4.5 & 7.4 & -0.003 (0.169) 	& 4.6 	& 4.4 \\
\rule{0pt}{1ex} \\

\multicolumn{9}{l}{\underline{3. Directional pleiotropy, InSIDE satisfied}} \\
\plei{}(0.01,0.004) &	& 0.043 (0.104) 	& 7.0 	& 0.099 (0.187)	& 5.8	& 8.0	& 0.001 (0.169)	& 5.9	& 4.8 \\
\plei{}(0.05,0.004)&	& 0.213 (0.105)		& 52.7	& 0.095 (0.187)	  & 33.3	& 7.6	& 0.000 (0.169)	& 37.2	& 4.5 \\
\plei{}(0.1,0.004) &	& 0.426 (0.107) 	& 96.3	& 0.096 (0.187)	
& 84.5	& 7.6	& -0.001 (0.169)	& 89.2	& 4.6 \\
\rule{0pt}{1ex} \\

\multicolumn{9}{l}{\underline{4. Directional pleiotropy, InSIDE violated}} \\
\plei{}(0.01,0.004) &	& 0.062 (0.104) 	& 9.5	& 0.184 (0.187)
& 4.6 	& 17.9	& 0.078 (0.169)	& 4.7	& 7.6 \\
\plei{}(0.05,0.004) &	& 0.235 (0.104)		& 62.1	& 0.187 (0.187)	
& 20.5 	& 18.3	& 0.082 (0.169)	& 22.3	& 7.5 \\
\plei{}(0.1,0.004) 	&	& 0.448 (0.106)		&97.9	& 0.181 (0.187)
& 73.3 	& 17.8	& 0.077 (0.169)	& 80.3	& 7.2 \\
\rule{0pt}{1ex} \\

\multicolumn{9}{l}{\textbf{Positive causal effect: \boldmath$\theta_{1}=0.3$}} \\
\multicolumn{9}{l}{\underline{1. No pleiotropy, InSIDE satisfied}} \\
				&	& 0.300 (0.047)		& 98.7	& 0.395 (0.158) & 4.4
& 70.8	& 0.299 (0.087)	& 3.9	& 86.2 \\
\rule{0pt}{1ex} \\

\multicolumn{9}{l}{\underline{2. Balanced pleiotropy, InSIDE satisfied}} \\
\plei{}(0,0.004) &	& 0.300 (0.104)		& 81.5	& 0.399 (0.187) & 4.4
& 58.0	& 0.301 (0.169)	& 4.6	& 44.4 \\
\rule{0pt}{1ex} \\

\multicolumn{9}{l}{\underline{3. Directional pleiotropy, InSIDE satisfied}} \\
\plei{}(0.01,0.004) &	& 0.342 (0.104)		& 89.4	& 0.395 (0.187)	
& 6.4	& 57.4	& 0.301 (0.169) & 5.9	& 44.4 \\
\plei{}(0.05,0.004) &	& 0.513 (0.105)		& 99.4	& 0.394	(0.187)	
& 33.0	& 57.4	& 0.296 (0.169)	& 38.0	& 43.4 \\
\plei{}(0.1,0.004) 	&	& 0.729 (0.107)		& 100.0	& 0.400 (0.187)	
& 83.5	& 58.2	& 0.304 (0.169)	& 88.6	& 45.5 \\
\rule{0pt}{1ex} \\

\multicolumn{9}{l}{\underline{4. Directional pleiotropy, InSIDE violated}} \\
\plei{}(0.01,0.004) &	& 0.365 (0.104)		& 92.1	& 0.489 (0.187)	
& 4.2	& 74.0	& 0.382 (0.169)	& 4.6	& 63.2 \\
\plei{}(0.05,0.004) &	 & 0.535 (0.104)	& 99.7	& 0.486 (0.187)	
& 20.3	& 72.9	& 0.382 (0.169)	& 21.1	& 63.2 \\
\plei{}(0.1,0.004) &	& 0.749 (0.106)		& 100.0	& 0.488 (0.187)	
& 72.5	& 73.4	& 0.381 (0.169)	& 79.6	& 62.8 \\
\bottomrule
\end{tabular}
}
\caption*{Abbreviations: MR, Mendelian randomization; SE, standard error; IVW, inverse-variance weighted; InSIDE, Instrument Strength Independent of Direct Effect.} \label{tab:cor}
\end{small} %
\end{center}
\end{table}

\pagebreak
\subsection*{Figures}

\clearpage
\renewcommand{\thesection}{A\arabic{section}}
\renewcommand{\thesubsection}{A.\arabic{subsection}}
\renewcommand{\thetable}{A\arabic{table}}
\renewcommand{\thefigure}{A\arabic{figure}}
\setcounter{table}{0}
\setcounter{figure}{0}
\setcounter{section}{0}
\setcounter{subsection}{0}
\setcounter{equation}{0}

\section*{Web Appendix}

\section{Sample software code}
\normalsize{We provide sample code written in R to perform the analyses described in this paper.  The associations of the genetic variants with the risk factors are denoted \texttt{bX\textit{k}} with standard error \texttt{bX\textit{k}se}, where $\texttt{\textit{k} = 1, \ldots, K}$. The associations of the genetic variants with the outcome are denoted \texttt{bY} with standard error \texttt{bYse}. The code for the multivariable models will be based on three risk factors and can be easily adapted to include the appropriate number of risk factors. It will be assumed that the causal effect of risk factor 1 on the outcome is of primary interest and all the genetic variants are uncorrelated.}

\subsubsection*{Inverse-variance weighted estimate:}
\normalsize{The inverse-variance weighted (IVW) estimate using summary statistics (equation~\ref{eq:IVW_simple}) can be calculated by:}

\scriptsize{
\begin{verbatim}
thetaUI     = sum(bY*bX1*bYse^-2)/sum(bX1^2*bYse^-2)
se_thetaUI  = 1/sqrt(sum(bX1^2*bYse^-2))
\end{verbatim}
}

\normalsize{The same IVW estimate using summary statistics can be obtained using weighted linear regression (equation~\ref{eq:IVW}):}

\scriptsize{
\begin{verbatim}
thetaUI            = summary(lm(bY~bX1-1, weights=bYse^-2))$coef[1]
se_thetaUI.fixed   = summary(lm(bY~bX1-1, weights=bYse^-2))$coef[1,2]/
                         summary(lm(bY~bX1-1, weights=bYse^-2))$sigma
se_thetaUI.random  = summary(lm(bY~bX1-1, weights=bYse^-2))$coef[1,2]/
                         min(summary(lm(bY~bX1-1, weights=bYse^-2))$sigma,1)
\end{verbatim}
}
\normalsize{In the fixed-effect model we divide the standard error of the causal estimate by the estimated residual standard error to force the residual standard error to be 1. For the multiplicative random-effect model the standard error is divided by the estimated residual standard error when the variability in the genetic associations is less than expected by chance (underdispersion). When there is evidence of heterogeneity between the causal estimates (overdispersion) the standard error is unaltered. The multiplicative random-effects model will result in a larger standard error compared to the fixed-effect model if there is heterogeneity between the causal estimates.}  The causal estimate obtained from the fixed- and multiplicative random-effects models will be the same.

\subsubsection*{Univariable MR-Egger:}
\normalsize{The univariable MR-Egger method is the same as the IVW method using weighted linear regression except the intercept term is estimated rather than being set to zero. Testing whether the intercept term is equal to zero is equivalent to testing for directional pleiotropy and the validity of the InSIDE assumption. The genetic associations with the risk factor \texttt{bX1} and outcome \texttt{bY} must be orientated with respect to the risk increasing or decreasing allele of the risk factor.  Under the MR-Egger model, multiplicative random-effects should be used as the presence of pleiotropy will lead to overdispersion. Since the residual standard error is estimated, we use the t-distribution with $J-2$ degrees of freedom for inference.}

\scriptsize{
\begin{verbatim}
#Orientation of the genetic associations
bY<-ifelse(bX1>0, bY, bY*-1)
bX1<-abs(bX1)
#Causal estimate
thetaUE            = summary(lm(bY~bX1, weights=bYse^-2))$coef[2]
se_thetaUE.random  = summary(lm(bY~bX1, weights=bYse^-2))$coef[2,2]/
                         min(summary(lm(bY~bX1, weights=bYse^-2))$sigma,1)
lb_thetaUE         = thetaUE - qt(0.975,df=length(bX1)-2)*se_thetaUE.random
ub_thetaUE         = thetaUE + qt(0.975,df=length(bX1)-2)*se_thetaUE.random
p_thetaUE          = 2*(1-pt(abs(thetaUE/se_thetaUE.random),df=length(bX1)-2))
#Test for directional pleiotropy
interUE            = summary(lm(bY~bX1, weights=bYse^-2))$coef[1]
se_interUE.random  = summary(lm(bY~bX1, weights=bYse^-2))$coef[1,2]/
                         min(summary(lm(bY~bX1, weights=bYse^-2))$sigma,1)
p_interUE          = 2*(1-pt(abs(interUE/se_interUE.random),df=length(bX1)-2))
\end{verbatim}
}

\subsubsection*{Multivariable IVW:}
\normalsize{The multivariable IVW method expands the IVW method using weighted linear regression by estimating the causal effects of the additional risk factors on the outcome.  We will include additional two risk factors and assume the causal estimate of interest is the effect of risk factor 1 on the outcome. Either fixed- or multiplicative random-effects can be used to estimate the standard error of the causal effect.

\scriptsize{
\begin{verbatim}
theta1MI            = summary(lm(bY~bX1+bX2+bX3-1, weights=bYse^-2))$coef[1]
se_theta1MI.fixed   = summary(lm(bY~bX1+bX2+bX3-1, weights=bYse^-2))$coef[1,2]/
                          summary(lm(bY~bX1+bX2+bX3-1, weights=bYse^-2))$sigma
se_theta1MI.random  = summary(lm(bY~bX1+bX2+bX3-1, weights=bYse^-2))$coef[1,2]/
                          min(summary(lm(bY~bX1+bX2+bX3-1, weights=bYse^-2))$sigma,1)
\end{verbatim}
}

\subsubsection*{Multivariable MR-Egger:}
\normalsize{The multivariable MR-Egger method is equivalent to the multivariable IVW method using weighted linear regression except the intercept is estimated rather than being set to zero. Testing whether the intercept term is equal to zero is equivalent to testing for directional pleiotropy and the validity of the InSIDE assumption. As with univariable MR-Egger, the standard errors should be calculated from the multiplicative random-effects model. The genetic associations should be orientated with respect to the risk increasing or decreasing allele of the risk factor of interest. In this sample code we will assume the causal effect of risk factor 1 is of primary interest. Since the residual standard error is estimated for the multivariable MR-Egger model we use the t-distribution with $J-(K+1)$ degrees of freedom for inference. }

\scriptsize{
\begin{verbatim}
#Orientation of the genetic associations with respect to X1
clist<-c("bX2","bX3","bY")
for (var in clist){
  eval(parse(text=paste0(var,"<-ifelse(bX1>0,",var,",",var,"*-1)")))
}
bX1<-abs(bX1)
#Causal estimate for X1
theta1ME            = summary(lm(bY~bX1+bX2+bX3, weights=bYse^-2))$coef[2]
se_theta1ME.random  = summary(lm(bY~bX1+bX2+bX3, weights=bYse^-2))$coef[2,2]/
                         min(summary(lm(bY~bX1+bX2+bX3, weights=bYse^-2))$sigma,1)
lb_theta1ME         = theta1ME - qt(0.975,df=length(bX1)-4)*se_theta1ME.random
ub_theta1ME         = theta1ME + qt(0.975,df=length(bX1)-4)*se_theta1ME.random
p_theta1ME          = 2*(1-pt(abs(theta1ME/se_theta1ME.random),df=length(bX1)-4))
#Test for directional pleiotropy
interME            = summary(lm(bY~bX1+bX2+bX3, weights=bYse^-2))$coef[1]
se_interME.random  = summary(lm(bY~bX1+bX2+bX3, weights=bYse^-2))$coef[1,2]/
                         min(summary(lm(bY~bX1+bX2+bX3, weights=bYse^-2))$sigma,1)
p_interME          = 2*(1-pt(abs(interME/se_interME.random),df=length(bX1)-4))
\end{verbatim}
}

\clearpage
\begin{bibunit}[wileyj]
\section{Comparison between the precision of the causal estimates from univariable and multivariable MR-Egger}
\normalsize{In this section, we compare the precision of the causal estimates from the univariable ($\hat{\theta}_{1UE}$) and multivariable ($\hat{\theta}_{1ME}$) MR-Egger models. For the multivariable model, we consider the genetic associations $\boldsymbol{\beta_{X_{k}}}$ with two risk factors ($k=2$), where the variance of the multivariable MR-Egger estimate $\hat{\theta}_{1ME}$ is given by:
\begin{align}
\var(\hat{\theta}_{1ME}) &= \frac{\phi^2\var(\boldsymbol{\beta_{X_{2}}})}{N(\var(\boldsymbol{\beta_{X_{1}}})\var(\boldsymbol{\beta_{X_{2}}})-\cov(\boldsymbol{\beta_{X_{1}}},\boldsymbol{\beta_{X_{2}}})^{2})} \notag \\
&\propto [\var(\boldsymbol{\beta_{X_{1}}})(1-\cor(\boldsymbol{\beta_{X_{1}}},\boldsymbol{\beta_{X_{2}}})^{2})]^{-1}
\end{align}

Under a fixed-effect model, the variance of the univariable MR-Egger estimate is proportional to the inverse of $\var(\boldsymbol{\beta_{X_{1}}})$ \cite{bowden2017pleio}. The estimate from the multivariable MR-Egger model $\hat{\theta}_{1ME}$ will be more precise than its univariable counterpart $\hat{\theta}_{1UE}$ if:
\begin{equation}
\frac{1}{\var(\boldsymbol{\beta_{X_{1}}})}>\frac{1}{\var(\boldsymbol{\beta_{X_{1}}})(1-\cor(\boldsymbol{\beta_{X_{1}}},\boldsymbol{\beta_{X_{2}}})^{2})}
\end{equation}
From the above inequality, $\hat{\theta}_{1UE}$ will always be more precise than $\hat{\theta}_{1ME}$ when $\boldsymbol{\beta_{X_{1}}}$ and $\boldsymbol{\beta_{X_{2}}}$ are correlated.  Under a multiplicative random-effects model (used throughout this paper), the variance of the residual error is estimated under the univariable MR-Egger model ($\phi_{UE}^{2}$) and the multivariable MR-Egger model ($\phi_{ME}^{2}$). For $\hat{\theta}_{1ME}$ to be more precise than $\hat{\theta}_{1UE}$, we require:
\begin{equation}
\frac{\phi_{UE}^{2}}{\var(\boldsymbol{\beta_{X_{1}}})}>\frac{\phi_{ME}^{2}}{\var(\boldsymbol{\beta_{X_{1}}})(1-\cor(\boldsymbol{\beta_{X_{1}}},\boldsymbol{\beta_{X_{2}}})^{2})}
\end{equation}
If $\boldsymbol{\beta_{X_{2}}}$ explains additional independent variability in the genetic associations with the outcome $\boldsymbol{\beta_{Y}}$, and $\boldsymbol{\beta_{X_{1}}}$ and $\boldsymbol{\beta_{X_{2}}}$ are independent, then the estimate from multivariable MR-Egger will be more precise than the estimate from univariable MR-Egger.  If $\boldsymbol{\beta_{X_{1}}}$ and $\boldsymbol{\beta_{X_{2}}}$ are correlated, then the precision of $\hat{\theta}_{1ME}$ will depend upon the strength of the correlation between $\boldsymbol{\beta_{X_{1}}}$ and $\boldsymbol{\beta_{X_{2}}}$, and the amount of additional independent variability $\boldsymbol{\beta_{X_{2}}}$ explains in $\boldsymbol{\beta_{Y}}$.  As the correlation between $\boldsymbol{\beta_{X_{1}}}$ and $\boldsymbol{\beta_{X_{2}}}$ increases, and $\boldsymbol{\beta_{X_{2}}}$ explains no additional independent variability in $\boldsymbol{\beta_{Y}}$, the precision of the multivariable MR-Egger estimate $\hat{\theta}_{1ME}$ will decrease.  
}

\section{Summary statistics for the data used in the applied example}
\normalsize{Applied MR studies usually report the proportion of variance in the exposure explained by the genetic variants ($R^{2}$) and a measurement of the strength of the instrumental variables (F-statistic). Instrumental variables are often considered to be strong if they have a F-statistic greater than 10 \cite{Burgess2011}.  Since the simulation study generated the summarized data directly, it was not possible to estimate the $R^{2}$ and F-statistic without making additional assumptions.  We therefore provide estimates of the $R^{2}$ and F-statistic from the applied example.  The F-statistic was calculated using the formula: 
\begin{equation}
\mbox{F-statisic}=\left(\frac{N-k-1}{k}\right)\left(\frac{R^2}{1-R^2}\right)
\end{equation}
where N is the number of participants (188,578) and k is the number of genetic variants (185).  For LDL-C, the estimated value for $R^{2}$ was 8.7\% and the F-statistic was 96.7.  For HDL-C, $R^{2}$ was 9.6\% and the F-statistic was 107.9, whilst triglycerides had a $R^{2}$ value of 5.9\% and F-statistic of 64.1. In this example, the F-statistics are large due to the large sample size and the high $R^{2}$ values.   
}
\clearpage

\section{Details and results from the simulation study investigating causal relationships between risk factors}
\normalsize{To investigate the behaviour of the multivariable MR-Egger method when causal relationships between risk factors exist, additional simulations were performed where $X_2$ was causally dependent on $X_1$. We assume that $X_2$ is causally dependent on $X_1$, and the causal effect of $X_1$ on $X_2$ is $\gamma$. The total causal effect of $X_1$ on $Y$ is $\theta_1 + \gamma\theta_2$; consisting of the direct effect ($\theta_1$) and the indirect effect via $X_2$ ($\gamma\theta_2$). The simulations outlined in Section 4 were repeated with the second line in the data generating model replaced with:
\begin{align}
\beta_{Y_{j}} &= \alpha'_{j} + \theta_{1}|\beta_{X_{1j}}| + \theta_{2}(\beta_{X_{2j}} + \gamma|\beta_{X_{1j}}|) + \theta_{3}\beta_{X_{3j}} +\epsilon_{j} 
\end{align}
The causal effect of $X_1$ on $X_2$ ($\gamma$) was set to 0.5. All other parameters were taken as in the original simulation study. $|\beta_{X_{1j}}|$, $(\beta_{X_{2j}} + \gamma|\beta_{X_{1j}}|)$, and $\beta_{X_{3j}}$ were the covariates included in the multivariable IVW and multivariable MR-Egger models. Note that the functional relationship between $X_1$ and $X_2$ induces a correlation structure between the covariates $|\beta_{X_{1j}}|$ and $(\beta_{X_{2j}} + \gamma|\beta_{X_{1j}}|)$ included in the multivariable models, even when $\boldsymbol{\beta_{X_{1}}}$ and $\boldsymbol{\beta_{X_{2}}}$ are generated independently. To account for the additional uncertainty in $\beta_{Y_{j}}$, the weights for univariable MR-Egger are given by equation \ref{wts: UV_2}, while the weights for multivariable IVW and multivariable MR-Egger were the same as the original simulation study (equation \ref{wts: MV}).  
\begin{align}
\se(\beta_{Y_{j}})^{-2}&={({\epsilon_{j}}^{2}+{\sigma_{\alpha'}}^{2}+{\theta_{2}}^{2}{\sigma_{2}}^{2} + {(\theta_2\gamma)}^2{\sigma_1}^2+2\theta_{2}\gamma\rho_{12}\sigma_{1}\sigma_{2} + {\theta_{3}}^{2}{\sigma_{3}}^{2})}^{-1} \label{wts: UV_2}
\end{align}

\subsection*{Results}
The results from the simulations that included a causal relationship between $X_1$ and $X_2$, using 10\thinspace000 simulated datasets, are presented in Web Table~\ref{tab:indep_2} ($\boldsymbol{\beta_{X_{k}}}$ generated independently, with the functional relationship between $X_1$ and $X_2$ inducing a correlation structure between $|\beta_{X_{1j}}|$ and $(\beta_{X_{2j}} + \gamma|\beta_{X_{1j}}|)$) and Web Table~\ref{tab:cor_2} ($\boldsymbol{\beta_{X_{k}}}$ correlated).

\textbf{$\boldsymbol{\beta_{X_{k}}}$ generated independently, with a correlation structure between the covariates $|\beta_{X_{1j}}|$ and $(\beta_{X_{2j}} + \gamma|\beta_{X_{1j}}|)$:} In scenarios where there was no bias in the original set of simulations, the multivariable IVW and multivariable MR-Egger methods consistently estimated the direct effect of $X_1$ on $Y$ ($\theta_1$), whilst the univariable MR-Egger method consistently estimated the total causal effect of $X_1$ on $Y$ ($\theta_1 + \gamma\theta_2$). Bias for the multivariable IVW method was present in scenarios 3 and 4 only, as in the original simulation study (Tables \ref{tab:indep} and \ref{tab:cor}). Compared to the results in Table \ref{tab:indep}, precision and power to detect a causal effect were reduced for the multivariable IVW and multivariable MR-Egger methods. This reduction in power may be due to the correlation structure between $|\beta_{X_{1j}}|$ and $(\beta_{X_{2j}} + \gamma|\beta_{X_{1j}}|)$, and the multivariable models conditioning on a mediator. Univariable and multivariable MR-Egger methods produced biased estimates of the total and direct causal effects in scenario 4 (InSIDE violated) only.  Unlike the original simulation study, precision and power to detect a causal effect were always better for the univariable MR-Egger method.  

\textbf{$\boldsymbol{\beta_{X_{k}}}$ correlated:} The multivariable IVW and multivariable MR-Egger methods estimated the direct effect of $X_1$ on $Y$, as in the independently generated setting. As with the original simulations (Tables \ref{tab:indep} and \ref{tab:cor}), the InSIDE assumption for univariable MR-Egger was violated for all four scenarios, resulting in biased point estimates.  However, as with the original simulation study, the multivariable InSIDE assumption was satisfied for scenarios 1,2 and 3, and so causal estimates from multivariable MR-Egger were unbiased. There was a more noticeable reduction in the precision and power to detect a causal effect for the multivariable IVW and multivariable MR-Egger methods under the correlated setting.  }

\clearpage 
\pagebreak
\begin{table}[htbp] 
\begin{center}
\begin{small}
\centering
\caption{Performance of multivariable IVW, univariable MR-Egger and multivariable MR-Egger with respect to $\hat{\theta}_{1}$ for a null ($\theta_{1}=0$) and positive ($\theta_{1}=0.3$) causal effect where $\boldsymbol{\beta_{X_{k}}}$ are generated independently for all $k$ (with a correlation structure between the covariates $|\beta_{X_{1j}}|$ and $(\beta_{X_{2j}} + \gamma|\beta_{X_{1j}}|)$), with a causal effect of $X_1$ on $X_2$ ($\gamma=0.5$). All tests were performed at the 5\% level of significance.}
\resizebox{\textwidth}{!} {
\begin{tabular}[c]{lR*{8}{c}}
\toprule
\multicolumn{2}{c}{} &
\multicolumn{2}{c}{\textbf{Multivariable IVW}} &
\multicolumn{3}{c}{\textbf{Univariable MR-Egger}} &
\multicolumn{3}{c}{\textbf{Multivariable MR-Egger}} \\
\cmidrule(r){3-4}\cmidrule(r){5-7}\cmidrule(r){8-10}
& & Mean $\hat{\theta}_{1}$ & Power, & Mean $\hat{\theta}_{1}$ & \multicolumn{2}{c}{Power, \%} & Mean $\hat{\theta}_{1}$ & \multicolumn{2}{c}{Power, \%} \\
& & (mean SE) & \% & (mean SE) & Intercept & Causal &
(mean SE) & Intercept & Causal \\
\toprule
\multicolumn{9}{l}{\textbf{Null causal effect: \boldmath$\theta_{1}=0$}} \\
\multicolumn{9}{l}{\underline{1. No pleiotropy, InSIDE satisfied}} \\
				&	& 0.000 (0.057) & 3.5 	& 0.051 (0.158) & 8.9 	& 5.8 & 0.001 (0.090) & 4.5 	& 4.2 \\
\rule{0pt}{1ex} \\ 	
							
\multicolumn{9}{l}{\underline{2. Balanced pleiotropy, InSIDE satisfied}} \\
\plei{}(0,0.004) &	& 0.001 (0.127) & 4.4 	& 0.049 (0.187) & 7.6 	& 5.6 & 0.001 (0.178) & 4.6 	& 4.2 \\
\rule{0pt}{1ex} \\

\multicolumn{9}{l}{\underline{3. Directional pleiotropy, InSIDE satisfied}} \\
\plei{}(0.01,0.004) &	& 0.041 (0.127) & 6.0 	& 0.049 (0.187) & 12.3 	& 5.4 & 0.000 (0.178) & 5.8  & 4.8 \\
\plei{}(0.05,0.004) &	& 0.195 (0.128) & 34.4 	& 0.048 (0.187) & 50.1 	& 5.3 & -0.001 (0.178) & 36.6 & 4.6 \\
\plei{}(0.1,0.004) 	&	& 0.393 (0.130) & 82.3 	& 0.052 (0.187) & 91.4 	& 5.6 & 0.002 (0.178) & 88.4 	& 4.7 \\
\rule{0pt}{1ex} \\

\multicolumn{9}{l}{\underline{4. Directional pleiotropy, InSIDE violated}} \\
\plei{}(0.01,0.004) &	& 0.076 (0.127) & 9.8 	& 0.138 (0.187) & 6.4  	& 11.6 & 0.088 (0.178) & 4.3  	& 7.6 \\
\plei{}(0.05,0.004) &	& 0.231 (0.127) & 45.2 	& 0.137 (0.187) & 34.4 	& 11.9 & 0.088 (0.178) & 21.7 	& 8.2 \\
\plei{}(0.1,0.004) 	&	& 0.426 (0.129) & 88.3 	& 0.141 (0.187) & 83.7 	& 11.9 & 0.089 (0.178) & 78.2 	& 8.1 \\
\rule{0pt}{1ex} \\

\multicolumn{9}{l}{\textbf{Positive causal effect: \boldmath$\theta_{1}=0.3$}} \\
\multicolumn{9}{l}{\underline{1. No pleiotropy, InSIDE satisfied}} \\
			&	& 0.301 (0.057) & 96.3 	& 0.353 (0.158) & 9.3 	& 62.3 & 0.301 (0.090) & 3.9 	& 84.6 \\
\rule{0pt}{1ex} \\

\multicolumn{9}{l}{\underline{2. Balanced pleiotropy, InSIDE satisfied}} \\
\plei{}(0,0.004) 	&	& 0.298 (0.127) & 65.4 	& 0.350 (0.187) & 7.4 	& 47.8 & 0.298 (0.178) & 4.4 	& 41.2 \\
\rule{0pt}{1ex} \\

\multicolumn{9}{l}{\underline{3. Directional pleiotropy, InSIDE satisfied}} \\
\plei{}(0.01,0.004) &	& 0.338 (0.127) & 75.5 	& 0.352 (0.187) & 11.8 	& 48.3 & 0.300 (0.178) & 6.1  & 41.1 \\
\plei{}(0.05,0.004) &	& 0.494 (0.128) & 95.2 	& 0.348 (0.188) & 49.2 	& 46.9 & 0.298 (0.179) & 36.8 & 40.3 \\
\plei{}(0.1,0.004)	&	& 0.689 (0.130) & 99.6 	& 0.347 (0.188) & 91.5 	& 47.1 & 0.296 (0.178) & 88.2 & 39.6	\\
\rule{0pt}{1ex} \\

\multicolumn{9}{l}{\underline{4. Directional pleiotropy, InSIDE violated}} \\
\plei{}(0.01,0.004) &	& 0.375 (0.127) & 82.6 	& 0.440 (0.187) & 6.6  	& 65.7 & 0.390 (0.178) & 4.7  & 60.1 \\
\plei{}(0.05,0.004) &	& 0.530 (0.128) & 97.0  & 0.438 (0.187) & 34.7	& 65.5 & 0.386 (0.178) & 21.7 & 59.9 \\
\plei{}(0.1,0.004) 	&	& 0.728 (0.129) & 99.7 	& 0.441 (0.187) & 83.6 	& 65.8 & 0.390 (0.178) & 78.5 & 60.1 \\
\bottomrule
\end{tabular}
}
\caption*{Abbreviations: MR, Mendelian randomization; SE, standard error; IVW, inverse-variance weighted; InSIDE, Instrument Strength Independent of Direct Effect.} \label{tab:indep_2}
\end{small} %
\end{center}
\end{table}

\pagebreak
\begin{table}[htbp] 
\begin{center}
\begin{small}
\centering
\caption{Performance of multivariable IVW, univariable MR-Egger and multivariable MR-Egger with $\boldsymbol{\beta_{X_{k}}}$ being correlated for all $k$, and a causal effect of $X_1$ on $X_2$}
\resizebox{\textwidth}{!} {
\begin{tabular}[c]{lR*{8}{c}}
\toprule
\multicolumn{2}{c}{} &
\multicolumn{2}{c}{\textbf{Multivariable IVW}} &
\multicolumn{3}{c}{\textbf{Univariable MR-Egger}} &
\multicolumn{3}{c}{\textbf{Multivariable MR-Egger}} \\
\cmidrule(r){3-4}\cmidrule(r){5-7}\cmidrule(r){8-10}
& & Mean $\hat{\theta}_{1}$ & Power, & Mean $\hat{\theta}_{1}$ & \multicolumn{2}{c}{Power, \%} & Mean $\hat{\theta}_{1}$ & \multicolumn{2}{c}{Power, \%} \\
& & (mean SE) & \% & (mean SE) & Intercept & Causal &
(mean SE) & Intercept & Causal \\
\toprule
\multicolumn{9}{l}{\textbf{Null causal effect: \boldmath$\theta_{1}=0$}} \\
\multicolumn{9}{l}{\underline{1. No pleiotropy, InSIDE satisfied}} \\
				&	& 0.000 (0.062) 	& 4.1 	& 0.146 (0.158) & 3.9 	& 15.6 & 0.000 (0.097) & 4.0 & 4.0 \\
\rule{0pt}{1ex} \\ 	
							
\multicolumn{9}{l}{\underline{2. Balanced pleiotropy, InSIDE satisfied}} \\
\plei{}(0,0.004) &		& 0.000 (0.137) & 4.5 	& 0.146 (0.188) & 4.1 	& 11.9 & 0.000 (0.190) & 4.6 	& 4.7 \\
\rule{0pt}{1ex} \\

\multicolumn{9}{l}{\underline{3. Directional pleiotropy, InSIDE satisfied}} \\
\plei{}(0.01,0.004) &	& 0.041 (0.137) & 5.7  	& 0.151 (0.187) & 5.4  	& 12.8 & 0.003 (0.189) & 5.7  & 4.4 \\
\plei{}(0.05,0.004) &	& 0.209 (0.138) & 34.2 	& 0.148 (0.187) & 32.8 	& 12.6 & 0.000 (0.190) & 36.9 & 4.7 \\
\plei{}(0.1,0.004) 	&	& 0.422 (0.140) & 82.2 	& 0.151 (0.188) & 83.0	& 12.9 & 0.004 (0.190) & 89.0 & 4.8 \\
\rule{0pt}{1ex} \\

\multicolumn{9}{l}{\underline{4. Directional pleiotropy, InSIDE violated}} \\
\plei{}(0.01,0.004) &	& 0.053 (0.137) & 6.2  	& 0.235 (0.188) & 4.3  	& 25.7 & 0.069 (0.189) & 4.9  & 6.4 \\
\plei{}(0.05,0.004) &	& 0.218 (0.137) & 37.2 	& 0.235 (0.188) & 20.3 	& 26.4 & 0.067 (0.189) & 21.8 & 6.7 \\
\plei{}(0.1,0.004) 	&	& 0.429 (0.139) & 84.3 	& 0.238 (0.188) & 71.3 	& 26.7 & 0.071 (0.189) & 79.2 & 6.6 \\
\rule{0pt}{1ex} \\

\multicolumn{9}{l}{\textbf{Positive causal effect: \boldmath$\theta_{1}=0.3$}} \\
\multicolumn{9}{l}{\underline{1. No pleiotropy, InSIDE satisfied}} \\
					&	& 0.299 (0.062) & 94.7 	& 0.446 (0.158) & 4.1 	& 79.7 & 0.300 (0.096) & 4.0 & 81.3 \\
\rule{0pt}{1ex} \\

\multicolumn{9}{l}{\underline{2. Balanced pleiotropy, InSIDE satisfied}} \\
\plei{}(0,0.004) 	&	& 0.301 (0.137) & 60.5 	& 0.445 (0.187) & 4.5 	& 66.6 & 0.300 (0.189) & 4.6 & 37.0 \\
\\
\rule{0pt}{1ex} \\

\multicolumn{9}{l}{\underline{3. Directional pleiotropy, InSIDE satisfied}} \\
\plei{}(0.01,0.004) &	& 0.339 (0.137) & 69.9	& 0.443 (0.188) & 5.7  & 66.1 & 0.296 (0.190) & 6.0  & 36.1 \\
\plei{}(0.05,0.004) &	& 0.510 (0.138) & 94.2	& 0.449 (0.188) & 32.6 & 67.7 & 0.302 (0.190) & 37.3 & 37.2 \\
\plei{}(0.1,0.004)	&	& 0.715 (0.140) & 99.2	& 0.445 (0.187) & 83.4 & 66.9 & 0.298 (0.189) & 89.4 & 36.8 \\
\rule{0pt}{1ex} \\

\multicolumn{9}{l}{\underline{4. Directional pleiotropy, InSIDE violated}} \\
\plei{}(0.01,0.004) &	& 0.353 (0.137) & 73.1	& 0.534 (0.188) & 4.4  & 79.4 & 0.367 (0.189) & 4.6  & 50.6 \\
\plei{}(0.05,0.004) &	& 0.519 (0.138) & 95.1 	& 0.534 (0.188) & 20.3 & 79.6 & 0.366 (0.190) & 21.7 & 50.5 \\
\plei{}(0.1,0.004) 	&	& 0.728 (0.139) & 99.5 	& 0.533 (0.188) & 72.5 & 79.6 & 0.368 (0.189) & 80.1 & 51.0 \\
\bottomrule
\end{tabular}
}
\caption*{Abbreviations: MR, Mendelian randomization; SE, standard error; IVW, inverse-variance weighted; InSIDE, Instrument Strength Independent of Direct Effect.} \label{tab:cor_2}
\end{small} %
\end{center}
\end{table}

\pagebreak
\section{Correlated genetic variants}
\normalsize{The methods discussed in this article have assumed that the genetic variants are uncorrelated (not in linkage disequilibrium).  There may, however, be cases where using multiple correlated variants from the same gene region will be more efficient than using uncorrelated variants from different gene regions \cite{burgess2016multiple}. If the genetic variants are in partial linkage disequilibrium, and each variant explains independent variation in the risk factor, then the inclusion of these variants will increase the power of the MR study. The precision of a MR study will not increase, however, if the variants are perfectly correlated.  

If correlated variants are included in an MR study, using summarized level data, the analysis should account for the correlation structure of the variants. If the correlation of the variants is not taken into consideration, the causal estimate will be too precise and this may lead to inappropriate inferences. To account for the correlation between the genetic variants for the univariable and multivariable IVW methods, we can use generalized weighted linear regression of the genetic associations, where the correlations of the variants are included in the weighting matrix, with the intercept set to zero \cite{burgess2015multivariable,burgess2016multiple}.  

If $\Omega_{st}=\se(\hat{\beta}_{Y_{s}})\se(\hat{\beta}_{Y_{t}})\rho_{st}$, where $\rho_{st}$ is the correlation between variants $s$ and $t$, then the causal estimate from a weighted generalised linear regression for univariable MR is:
\begin{equation}
\hat{\theta}_{UIC}=(\hat{\beta}^{T}_{X_{j}}\Omega^{-1}\hat{\beta}_{X_{j}})^{-1}\hat{\beta}^{T}_{X_{j}}\Omega^{-1}\hat{\beta}_{Y_{j}}
\end{equation}
with the standard error of the causal estimate: 
\begin{equation}
\hat{\theta}_{UIC}=\sqrt{(\hat{\beta}^{T}_{X_{j}}\Omega^{-1}\hat{\beta}_{X_{j}})^{-1}}
\end{equation}
Whilst the univariable MR-Egger estimates can be obtained by fitting the same generalized weighted linear regression model, but allowing the intercept term to be estimated, the effect of using correlated genetic variants in the univariable MR-Egger method has not been considered in detail. Further investigation into the impact correlated variants may have on the interpretation of the direct effect, and the InSIDE assumption, must be considered at the univariable level first, and then expanded to multivariable MR-Egger.} 
\clearpage
\putbib[ref_reduced]
\end{bibunit}
\end{document}